\documentclass[onecolumn,showpacs,preprintnumbers,elsart]{revtex4}
\usepackage{makeidx}
\usepackage{amssymb}
\usepackage{amsmath}
\usepackage{mathrsfs}
\usepackage{graphicx}
\usepackage{dcolumn}
\usepackage{bm}
\usepackage[center]{subfigure}
\usepackage{color}

\begin{document}

\title{Lattice solitons with quadrupolar intersite interactions}
\author{Yongyao Li$^{1,2,3}$, Jingfeng Liu$^{1}$, Wei Pang$^{4}$, and Boris
A. Malomed$^{2}$}
\email{malomed@post.tau.ac.il}
\affiliation{$^{1}$Department of Applied Physics, South China Agricultural University,
Guangzhou 510642, China \\
$^{2}$Department of Physical Electronics, School of Electrical Engineering,
Faculty of Engineering, Tel Aviv University, Tel Aviv 69978, Israel\\
$^{3}$Modern Educational Technology Center, South China Agricultural
University, Guangzhou 510642, China\\
$^{4}$Department of Experiment Teaching, Guangdong University of Technology,
Guangzhou 510006, China.}

\begin{abstract}
We study two-dimensional (2D) solitons in the mean-field models of ultracold
gases with long-range quadrupole-quadrupole interaction (QQI) between
particles. The condensate is loaded into a deep optical-lattice (OL)
potential, therefore the model is based on the 2D discrete nonlinear Schr%
\"{o}dinger equation with contact onsite and long-range intersite
interactions, which represent the QQI. The quadrupoles are built as pairs of
electric dipoles and anti-dipoles orientated perpendicular to the 2D plane
to which the gas is confined. Because the quadrupoles interact with the
local gradient of the external field, they are polarized by inhomogeneous dc
electric field that may be supplied by a tapered capacitor. Shapes,
stability, mobility, and collisions of fundamental discrete solitons are
studied by means of systematic simulations. In particular, threshold values
of the norm, necessary for the existence of the solitons, are found, and
anisotropy of their static and dynamical properties is explored. As concerns
the mobility and collisions, it is the first analysis of such properties for
discrete solitons on 2D lattices with long-range intersite interactions of
any type. Estimates demonstrate that the setting can be realized under
experimentally available conditions, predicting solitons built of $\sim
10,000$ particles.
\end{abstract}

\maketitle



\section{Introduction}

Interactions between particles play a crucial role in dynamics of
Bose-Einstein condensates (BECs). Ubiquitous are short-range contact
interactions, which are described by the single parameter, the \textit{s}%
-wave scattering length $a_{s}$ \cite{GPE,Courteille1,Inouye2}. More
specific are isotropic long-range Van der Waals interactions between Rydberg
atoms in ultracold bosonic gases \cite{Heidemann3,Viteau4}, and anisotropic
long-range dipole-dipole interactions (DDIs) in dipolar condensates \cite%
{Pfau}-\cite{Pollack19}, \cite{review7}. DDIs occur in gases formed by
magnetic atoms, such as Cr, Dy, Er \cite{Griesmaier23,LuM24,Aikawa25}, or
gases of molecules carrying electric dipole moments---for instance, CO, ND,
and OH \cite{Bethlem26,Hendrick27,Bochinski28}. Using external dc magnetic
or electric fields polarizing permanent atomic or molecular moments, a
plenty of controllable structures, including solitons, have been predicted
and demonstrated in dipolar condensates \cite{Pedri5,Tikhonenkov6,review7},
\cite{Gershon}-\cite{Arnaldo}, \cite{review8}. Many such structures are
supported by optical lattices (OLs). In the limit of deep OL\ potentials,
the description of the dipolar BEC is provided by discrete models with
long-range intersite interactions \cite%
{Belgrade,Zhihuan,nonpoly,Gligoric34,Goran,mix-demix,BuBu,Maluckov11,Peierls}%
. Moreover, it was predicted \cite{JS,HS} and demonstrated experimentally
\cite{JS2} that the interaction of particles carrying permanent electric
dipole moments with a singular dc field created by a charged wire or a
point-like charge may give rise both to the specific collapse mechanism, and
to its suppression by inter-particle interactions. In addition, the
repulsive DDIs between moments induced by nonuniform dc fields in a gas of
non-dipolar polarizable particles may give rise to bright solitons of a
completely different type \cite{we}.

In this work, we propose a possibility of the formation of two-dimensional
(2D) matter-wave solitons in non-dipolar molecular ultracold gas, loaded
into a deep OL potential, with long-range quadrupole-quadrupole interactions
(QQIs) between particles. Experimental methods for the measurement of
molecular quadrupole moments are well elaborated \cite{experiment}. It has
been reported that some simple molecules, such as acetylene \cite{C2H2},
have relatively large quadrupole moments ($\simeq 6$ Debye $\cdot \overset{%
\circ }{\mathrm{A}}$). Still larger moments were measured in long-lived
metastable states of alkaline-earth \cite{alkaline-earth,alkaline-earth2}
and ytterbium \cite{Yb} atoms ($\simeq 20$ Debye $\cdot \overset{\circ }{%
\mathrm{A}}$), and in alkaline diatoms \cite{diatoms} (up to $\simeq 50$
Debye $\cdot \overset{\circ }{\mathrm{A}}$). As recently shown for fermions,
such sizes of the quadrupole moments are sufficient to generate new phases
in quantum gases trapped in the OL \cite{Bhongale}. Furthermore, it has been
predicted that the QQI can be strongly amplified by means of optical
vortices \cite{vortex}.

\begin{figure}[tbp]
\centering\subfigure[] {\label{fig1a}
\includegraphics[scale=0.4]{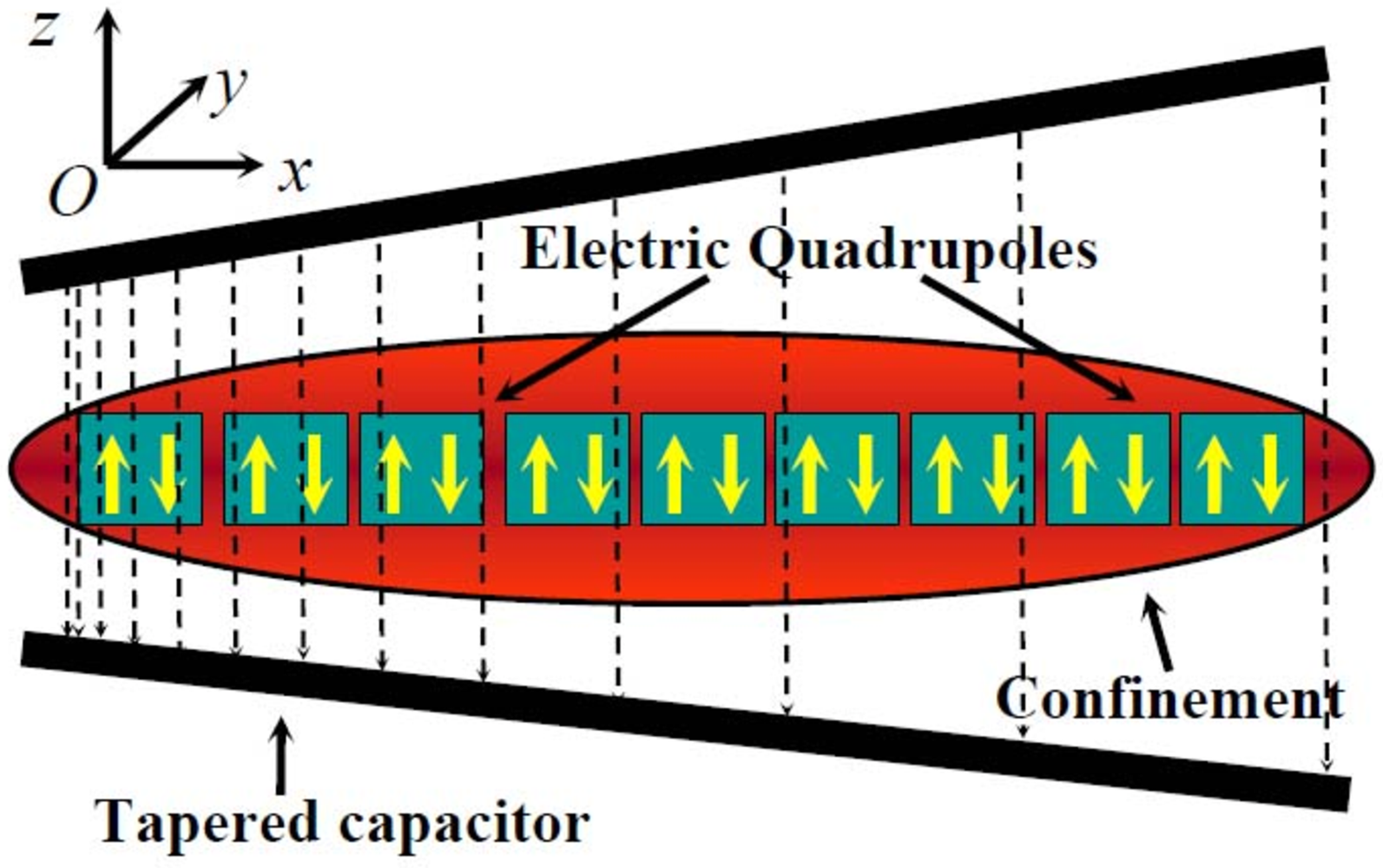}}%
\subfigure[] {\label{fig1b}
\includegraphics[scale=0.43]{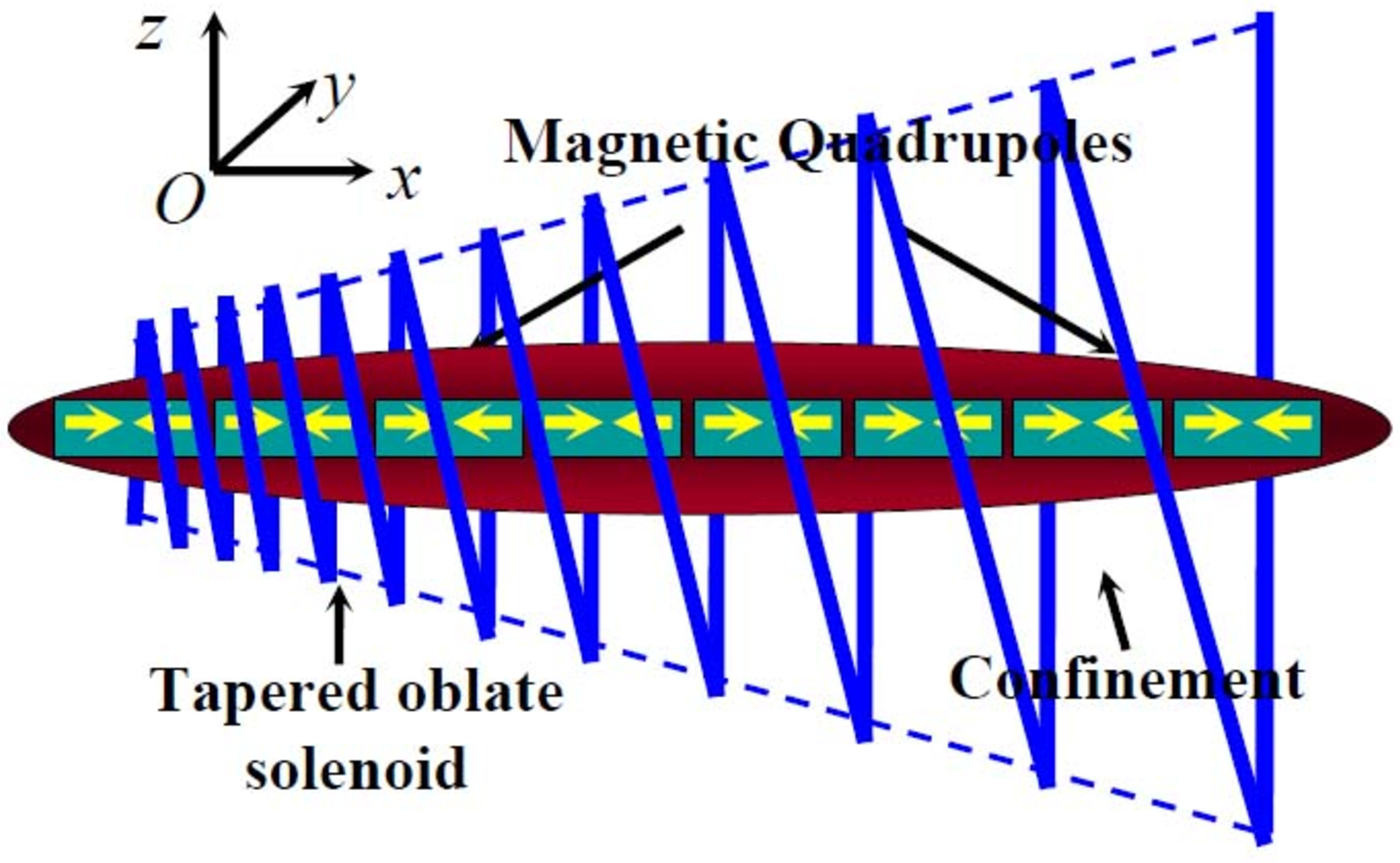}}
\caption{(Color online) (a) Electric quadrupoles, built as tightly bound
pairs of dipole and anti-dipoles directed perpendicular to the $(x,y)$
plane. They are polarized by the dc electric field which is also directed
along $z$, whose strength is a linear function of $x$. As shown here, the
field can be imposed by a tapered capacitor. (b) Magnetic quadrupoles, built
as tightly bound dipole-antidipole pairs directed along axis $x$.They can be
aligned by means of the dc magnetic field directed along $x$, whose strength
is a linear function of $x$. The magnetic field can be created by the
tapered solenoid, as shown in panel (b). }
\label{Fig1}
\end{figure}

There are two straightforward settings, for electric and magnetic
quadrupoles, respectively, which can feature QQIs in the 2D geometry. One
relies upon electric quadrupoles, built as tightly bound pairs of dipoles
and antidipoles, which are directed \emph{perpendicular} to the system's
plane, $\left( x,y\right) $, i.e., along axis $z$. Because the quadrupole
interacts with the gradient of the external field, rather than with the
field itself, in this setting the quadrupoles may be aligned (polarized) by
an external electric field which is also directed along $z$, and whose
strength is a linearly growing function of $x$. Such an electric field can
be imposed, in turn, by a \textit{tapered capacitor}, see Fig. \ref{fig1a}.

Quadrupoles (both electric and magnetic ones) are described by symmetric
traceless tensor $Q_{\alpha \beta }$ ($\alpha ,\beta =x,y,z$), with $%
\sum_{\alpha }Q_{\alpha \alpha }\equiv 0$ \cite{LL30}. In this notation, the
electric quadrupole defined above [in Fig. \ref{Fig1}(a)] has the following
components:%
\begin{equation}
Q_{xy}=Q_{yx}\equiv Q;~Q_{zz}=Q_{zx}=Q_{zx}=Q_{yz}=0,  \label{Q}
\end{equation}%
where the quadrupole moment per se is defined as
\begin{equation}
Q=3d\varepsilon ,  \label{lim}
\end{equation}%
with$\pm d$ and $\varepsilon $ being, respectively, the dipolar moments of
the bound dipoles and antidipole, and the distance between them.

The potential of the interaction between two quadrupoles of the present type
in the planar configuration considered here can be derived from the general
formula for the QQI potential \cite{formula}, the result being%
\begin{equation}
U_{\mathrm{QQ}}^{\mathrm{(electr)}}\left( r,\theta \right) =\frac{4}{3}%
Q^{2}r^{-5}\left( 1-5\cos ^{2}\theta \right) ,  \label{electrpot}
\end{equation}%
where $r$ is the distance between the quadrupoles, and $\theta $ the angle
between the vector connecting the quadrupoles and the line connecting the
dipole and antidipole inside the quadrupole (the latter direction is here
defined as axis $x$).

On the other hand, it is possible to introduce magnetic quadrupoles, built
as tightly bound pairs of dipoles and antidipoles directed \emph{in-plane},
along axis $x$. Because these quadrupoles too interact with the gradient of
the external field, they may be aligned by means of the magnetic field
directed along $x$, whose strength grows linearly with $x$. This field may
be applied by a \textit{tapered} \textit{oblate solenoid}, see Fig. \ref%
{fig1b}. The respective quadrupole tensor is
\begin{equation}
Q_{yy}=Q_{zz}=-\frac{1}{2}Q_{xx},~Q_{xy}=Q_{zx}=Q_{yz}=0.  \label{standard}
\end{equation}%
The potential of the interaction between the quadrupoles of the present type
can also be derived from the general formula \cite{formula}, yielding
\begin{equation}
U_{\mathrm{QQ}}^{\mathrm{(magn)}}\left( r,\theta \right) =\frac{4}{3}%
Q^{2}r^{-5}\left( 3-30\cos ^{2}\theta +35\cos ^{4}\theta \right) .
\label{magnpot}
\end{equation}

Averaging potentials (\ref{electrpot}) and (\ref{magnpot}) over the entire
angular range, $0\leq \theta <2\pi $, yields mean values which corresponds
to attraction and repulsion, respectively:
\begin{equation}
\frac{1}{2\pi }\int_{0}^{2\pi }U_{\mathrm{QQ}}^{\mathrm{(electr)}}\left(
R,\theta \right) d\theta =-2Q^{2}r^{-5}<0.  \label{average}
\end{equation}%
\begin{equation}
\frac{1}{2\pi }\int_{0}^{2\pi }U_{\mathrm{QQ}}^{\mathrm{(magn)}}\left(
R,\theta \right) d\theta =\frac{3}{2}Q^{2}r^{-5}>0,
\end{equation}%
hence potential (\ref{electrpot}) has a chance to create 2D solitons, while
for potential (\ref{magnpot}) this is not plausible. Indeed, numerical tests
with interaction kernel (\ref{magnpot}) have not revealed soliton modes.
Therefore, in what follows below we consider the model with the QQI defined
as per Eq. (\ref{electrpot}) and Fig. \ref{fig1a}.

For the sake of comparison, it is relevant to compare expression (\ref%
{electrpot}) and (\ref{magnpot}) with the commonly known potential of the
DDI for a pair of in-plane-oriented parallel dipoles with moment $D$:%
\begin{equation}
U_{\mathrm{DD}}\left( r,\theta \right) =\frac{1}{3}D^{2}r^{-3}\left( 1-3\cos
^{2}\theta \right) .  \label{DD}
\end{equation}%
Note that this potential is attractive, on the average:%
\begin{equation}
\frac{1}{2\pi }\int_{0}^{2\pi }U_{\mathrm{DD}}\left( R,\theta \right)
d\theta =-\frac{1}{6}D^{2}r^{-3}<0.
\end{equation}%
Accordingly, the existence of stable anisotropic 2D solitons in this setting
was first demonstrated in Ref. \cite{Tikhonenkov6}.

The objective of this work is to study shapes, stability, mobility and
collisions of 2D matter-wave solitons in BEC formed by quadrupole particles
trapped in a deep optical lattice (OL). The discrete model describing the
present setting is derived in Sec. II, where we also present estimates of
physical parameters of the setting, and of the expected soliton states.
Results of systematic numerical studies of static 2D solitons in the model
are reported in Sec. III, and their dynamical properties, namely, mobility
and collisions, are presented in Section IV. It is relevant to stress that
mobility and collisions of 2D discrete solitons were not previously studied
in any lattice system with long-range interactions. The paper is concluded
by Sec. V.

\section{The model equation}

The underlying 3D Gross-Pitaevskii equation (GPE) \cite{GPE}, which takes
into regard both the contact isotropic nonlinearity and the long-range QQI
corresponding to potential (\ref{electrpot}), can be reduced to the
normalized 2D equation,
\begin{gather}
i\frac{\partial \psi }{\partial t}+\frac{1}{2}\left( \frac{\partial ^{2}}{%
\partial x^{2}}+\frac{\partial ^{2}}{\partial y^{2}}\right) \psi
-g\left\vert \psi \right\vert ^{2}\psi -U\left( x,y\right) \psi   \notag \\
-G\psi (\mathbf{r})\int \left\vert \psi \left( \mathbf{r}^{\prime }\right)
\right\vert ^{2}\left( 1-5\cos ^{2}\theta \right) \frac{d\mathbf{r}^{\prime }%
}{\left\vert \mathbf{r}-\mathbf{r}^{\prime }\right\vert ^{5}}=0,  \label{GP}
\end{gather}%
where $\mathbf{r}\equiv \left\{ x,y\right\} $, angle $\theta $ has the same
meaning as in Eq. (\ref{electrpot}), and $U\left( x,y\right) $ represents
the OL potential. The derivation of Eq. (\ref{GP}) assumes the factorization
of the 3D\ mean-field wave function, $\Psi $, under the action of the tight
trapping potential imposed in the transverse direction:%
\begin{equation}
\Psi \left( X,Y,Z,T\right) =\left( \sqrt{\pi }a_{\perp }^{3}\right)
^{-1/2}\exp \left( -\frac{i\hbar }{2ma_{\perp }^{2}}t-\frac{1}{2}%
z^{2}\right) \psi \left( x,y,t\right) ,  \label{Psi}
\end{equation}%
where the scaled coordinates and time are related to physical ones, $X,Y,Z,T$%
, as follows: $\left\{ X,Y,Z\right\} \equiv a_{\perp }\left\{ x,y,z\right\}
,~T\equiv \left( ma_{\perp }^{2}/\hbar \right) t,~a_{\perp }$ is the
transverse localization length, and $m$ is the mass of the particle. As it
follows from factorized ansatz (\ref{Psi}), the total number of particles in
the condensate is given by%
\begin{equation}
P_{\mathrm{total}}=\int \int \left\vert \psi \left( x,y\right) \right\vert
^{2}dxdy.  \label{N}
\end{equation}%
The rescaling implied above gives rise to expressions for dimensionless
coefficients of the contact and long-range interactions in Eq. (\ref{GP}) in
terms of the scattering length, $a_{s}$, and quadrupole moment:
\begin{equation}
g=2\sqrt{2\pi }\frac{a_{s}}{a_{\perp }},~~G=\frac{4mQ^{2}}{3\hbar
^{2}a_{\perp }^{3}}.  \label{g-G}
\end{equation}

Equation (\ref{GP}) belongs to the class of nonlocal nonlinear Schr\"{o}%
dinger (NLS) equations, which, unlike their local counterpart with the
self-attractive cubic term, are free of the collapse in the 2D geometry \cite%
{PRE2002,review2004}, therefore 2D solitons are stable in such models, on
the contrary to the commonly known instability of Townes solitons in the
local 2D NLS equation \cite{PRE2006}. Further, the nonlocality affects the
character of the interaction between 2D solitons. In particular,
interactions between dark solitons, and between out-of-phase bright modes,
can be made attractive, in contrast with the repulsion in local models, see
a brief review \cite{review2004} for nonlocal NLS equations in models of
nonlinear optics.

The next step is to replace continuous equation (\ref{GP}) by its discrete
counterpart, corresponding to the condensate fragmented by a deep OL
potential. To this end, the continuous wave function is decomposed over a
set of modes strongly localized in vicinity of each OL site (Wannier
functions),%
\begin{equation}
\psi \left( x,y,t\right) =\sum_{m,n}\psi _{mn}(t)\Phi _{mn}(x,y),
\end{equation}%
where $m,n$ are discrete coordinates on the lattice, as it was done in the
course of the derivation of the 1D \cite{Belgrade} and 2D \cite{Gligoric34}
discrete models for the dipolar BEC\ trapped in the deep OL\ potentials. The
result is the following rescaled discrete version of Eq. (\ref{GP}), i.e., a
2D discrete NLS equation with long-range interactions:%
\begin{gather}
\partial _{t}\psi _{mn}=-\frac{1}{2}\left( \psi _{m+1,n}+\psi _{m-1,n}+\psi
_{m,n+1}+\psi _{m,n-1}-4\psi _{mn}\right) +\sigma |\psi _{mn}|^{2}\psi _{mn}
\notag \\
+\psi _{mn}\sum_{\left\{ m^{\prime },n^{\prime }\right\} \neq \left\{
m,n\right\} }f_{\mathrm{QQ}}(m-m^{\prime },n-n^{\prime })|\psi _{m^{\prime
}n^{\prime }}|^{2},  \label{discr}
\end{gather}%
with the discrete QQI kernel,
\begin{equation}
f_{\mathrm{QQ}}(m-m^{\prime },n-n^{\prime })={\frac{(n-n^{\prime
})^{2}-4(m-m^{\prime })^{2}}{[(m-m^{\prime })^{2}+(n-n^{\prime })^{2}]^{7/2}}%
}.  \label{QQInonlocal}
\end{equation}%
The rescaling is performed here in the same way as it was done in Ref. \cite%
{Gligoric34}, using coefficients which are expressed in the form of
normalization and overlapping integrals built of the Wannier functions. Note
that the coefficient in front of the QQI terms is scaled here to be $1$,
while $\sigma $ is the relative strength of the contact interactions.
Obviously, Eq. (\ref{discr}) conserves the norm of the discrete wave
function, which is the discrete counterpart of norm (\ref{N}):
\begin{equation}
P=\sum_{m,n}|\psi _{mn}|^{2},  \label{P}
\end{equation}

It is well known that the discretization of the NLS equation is another
general mechanism helping to suppress the collapse and ensuing instability
of solitons in the 2D\ NLS equation \cite{nonlinearity1994}. Likewise, the
discreteness readily stabilizes 2D discrete solitons with embedded vorticity
\cite{discr-vort}. Furthermore, it was demonstrated, both theoretically and
experimentally, that stable 2D gap solitons are supported by a system
modeled by the discrete NLS equation with long-range inter-site interactions
\cite{gap-sol}. It is relevant to mention too that, in addition to its
realizations in optics and BEC, the discrete NLS equation with long-range
intersite interactions models the so-called Scheibe aggregates of closely
packed molecules, in which soliton solutions are known as well \cite{Scheibe}%
.

The analysis reported below is based on Eq. (\ref{discr}). Note that the
discrete equation which was derived in Refs. \cite{Gligoric34,Goran} for the
dipolar BEC trapped in the deep OL potential differs by the form of the
interaction kernel, which is the discrete counterpart of the DDI kernel (\ref%
{DD}):
\begin{equation}
f_{\mathrm{DD}}(m-m^{\prime },n-n^{\prime })={\frac{(n-n^{\prime
})^{2}-2(m-m^{\prime })^{2}}{[(m-m^{\prime })^{2}+(n-n^{\prime })^{2}]^{5/2}}%
}.  \label{DDInonlocal}
\end{equation}%
It is also relevant to mention that Eq. (\ref{discr}) with $\sigma =0$
applies as well to fermion lattice gases with the long-range interaction,
cf. Ref. \cite{Bhongale}.

Finally, to estimate the range of physical parameters allowing the
implementation of the present model, the magnitude of the quadrupole moment
can be estimated for a complex built as a square with charges $\left(
+e,-e,+e,-e\right) $ set at its vertices, with linear size $\varepsilon $ ($%
e $ is the electron's charge). According to Eq. (\ref{lim}), the
corresponding quadrupole momentum is
\begin{equation}
Q=3e\varepsilon ^{2}.  \label{Q3}
\end{equation}%
Then, adopting typical values $a_{s}\simeq 5$ nm, $a_{\perp }\simeq 1$ $%
\mathrm{\mu }$m, and $m$ $\simeq 100$ proton masses, Eqs. (\ref{GP}), (\ref%
{g-G}), and (\ref{Q3}) can be used to estimate the linear size of the
quadrupole, $\varepsilon $, which is necessary to make the strength of the
QQI comparable to that of the contact interactions. The result is $%
\varepsilon \sim 1~\mathrm{nm}$, which is quite realistic for small
molecules. Further, typical scaled characteristics of solitons reported
below, if translated back into physical units, lead to an estimate $\sim 30$
$\mathrm{\mu m}$ for the linear size of 2D solitons built of $\sim 10,000$
particles.

\section{Numerical results}

\subsection{The shape and stability of the solitons}

Stationary solutions to Eq. (\ref{discr}) with real chemical potential $\mu $
were looked for in the usual form,
\begin{equation}
\psi _{mn}(t)=\phi _{mn}e^{-i\mu t}.  \label{mu}
\end{equation}%
Stationary profiles $\phi _{mn}$ were found, in a finite domain of size $%
N\times N$, by means of the well-known imaginary-time propagation method
\cite{Chiofalo31,JYang32,JYang33}. The stability of the so found solitons
was then tested by simulations of their evolution in real time, adding
random perturbations to the initial conditions. It is instructive to present
the results for solitons maintained by the QQI, comparing them to those
obtained in the model with the DDI kernel (\ref{DD}) in Refs. \cite%
{Gligoric34} and \cite{Goran}.

For both models, QQI and DDI, the numerical results reveal a critical value
of the norm, $P_{\mathrm{cr}}$, below which discrete wave packets do not
self-trap into 2D solitons, but rather spread out into almost flat states
via the delocalization transition, which is a generic feature of 2D systems
\cite{BBB}. In fact, $P_{\mathrm{cr}}$ originates from the norm of the
\textit{Townes soliton}, which is the single value at which the degenerate
family of solitons exist in the uniform 2D medium with the self-attractive
local cubic nonlinearity \cite{VK1}, and which is stretched into a finite
existence interval under the stabilizing action of the OL potential \cite%
{BBB0}. We have found that all the solitons existing at $P>P_{\mathrm{cr}}$
are stable.

\begin{figure}[tbp]
\centering\subfigure[] {\label{fig2a}
\includegraphics[scale=0.25]{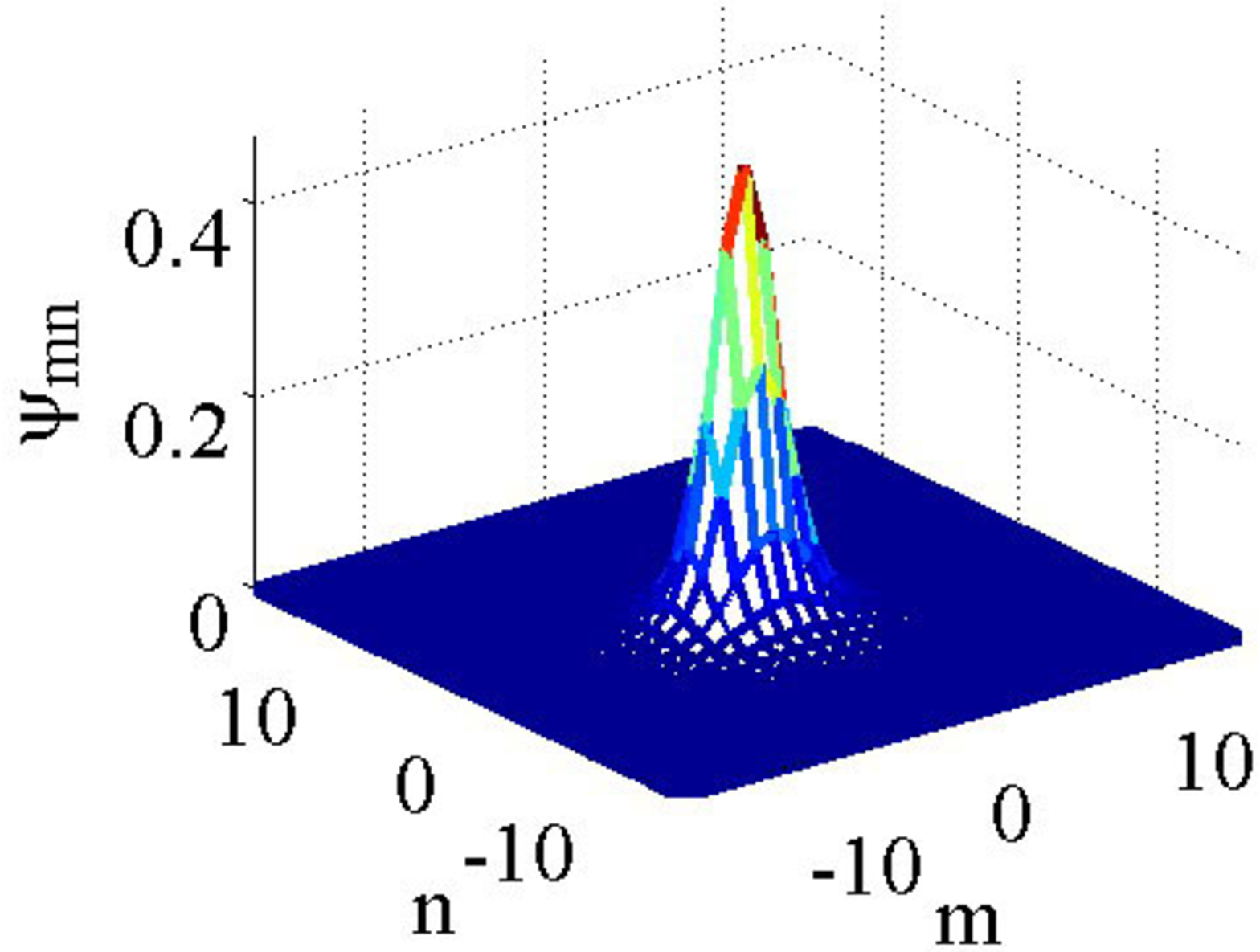}}%
\subfigure[] {\label{fig2b}
\includegraphics[scale=0.25]{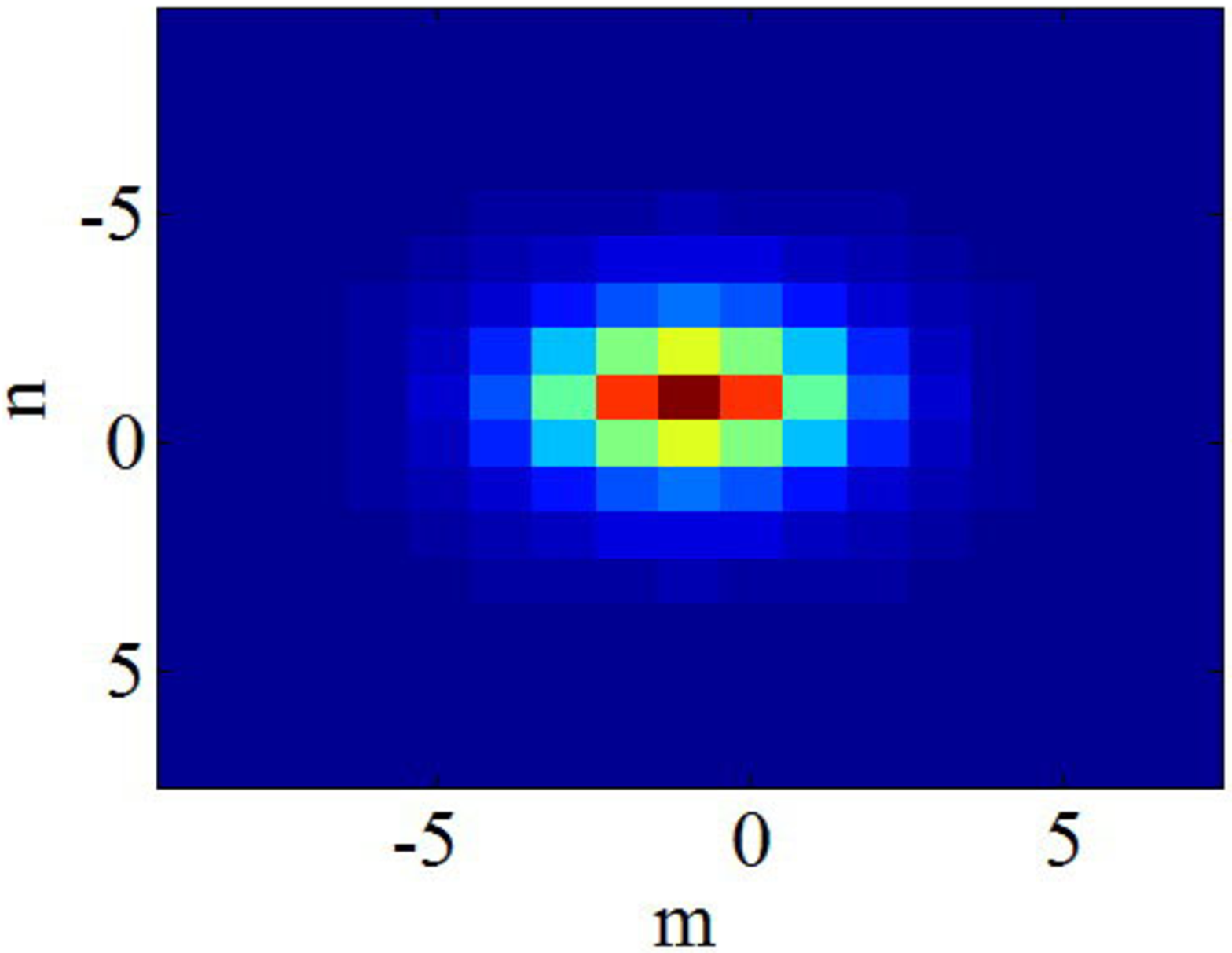}}
\caption{(Color online) Side (a) and top (b) views of a stable discrete
soliton supported by the long-range quadrupole interactions. The soliton was
found as a stationary solution of Eq. (\protect\ref{discr}) with $\protect%
\sigma =0$ (no contact interactions) and norm $P=1.2$, on the lattice of
size $32\times 32$. In this case, $P_{cr}=0.86$, see Fig. \protect\ref{fig3b}%
.}
\label{examplessolution}
\end{figure}

A typical example of a discrete soliton \ maintained by the QQI nonlinearity
is displayed in Fig. \ref{examplessolution}. As expected, the soliton's
shape is anisotropic, being prolate in the $m$ direction, in accordance with
the fact that the QQI kernel (\ref{QQInonlocal}) is attractive along $m$ and
repulsive in the perpendicular direction.

\begin{figure}[tbp]
\centering\subfigure[] {\label{fig3a}
\includegraphics[scale=0.22]{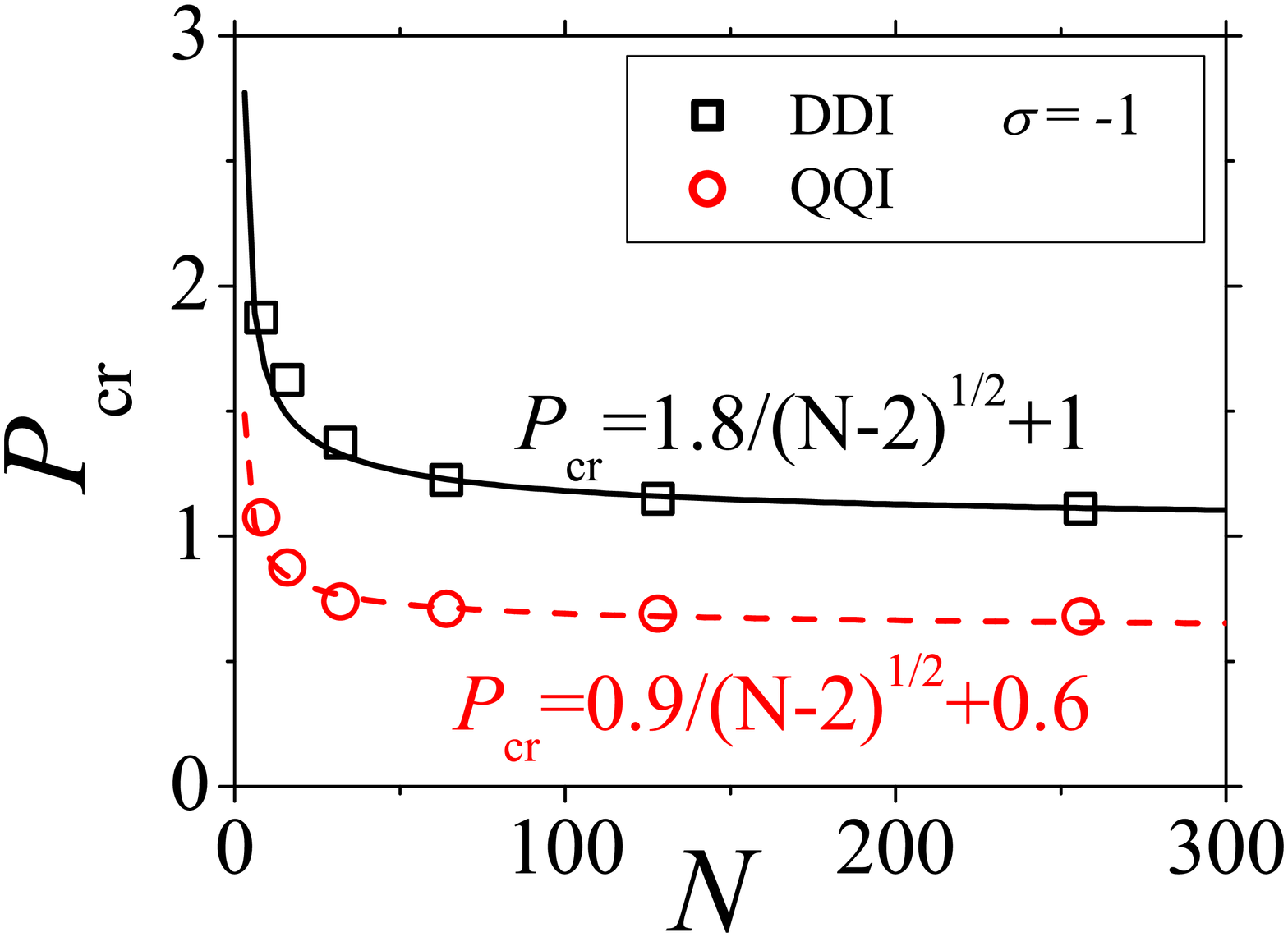}}%
\subfigure[] {\label{fig3b}
\includegraphics[scale=0.22]{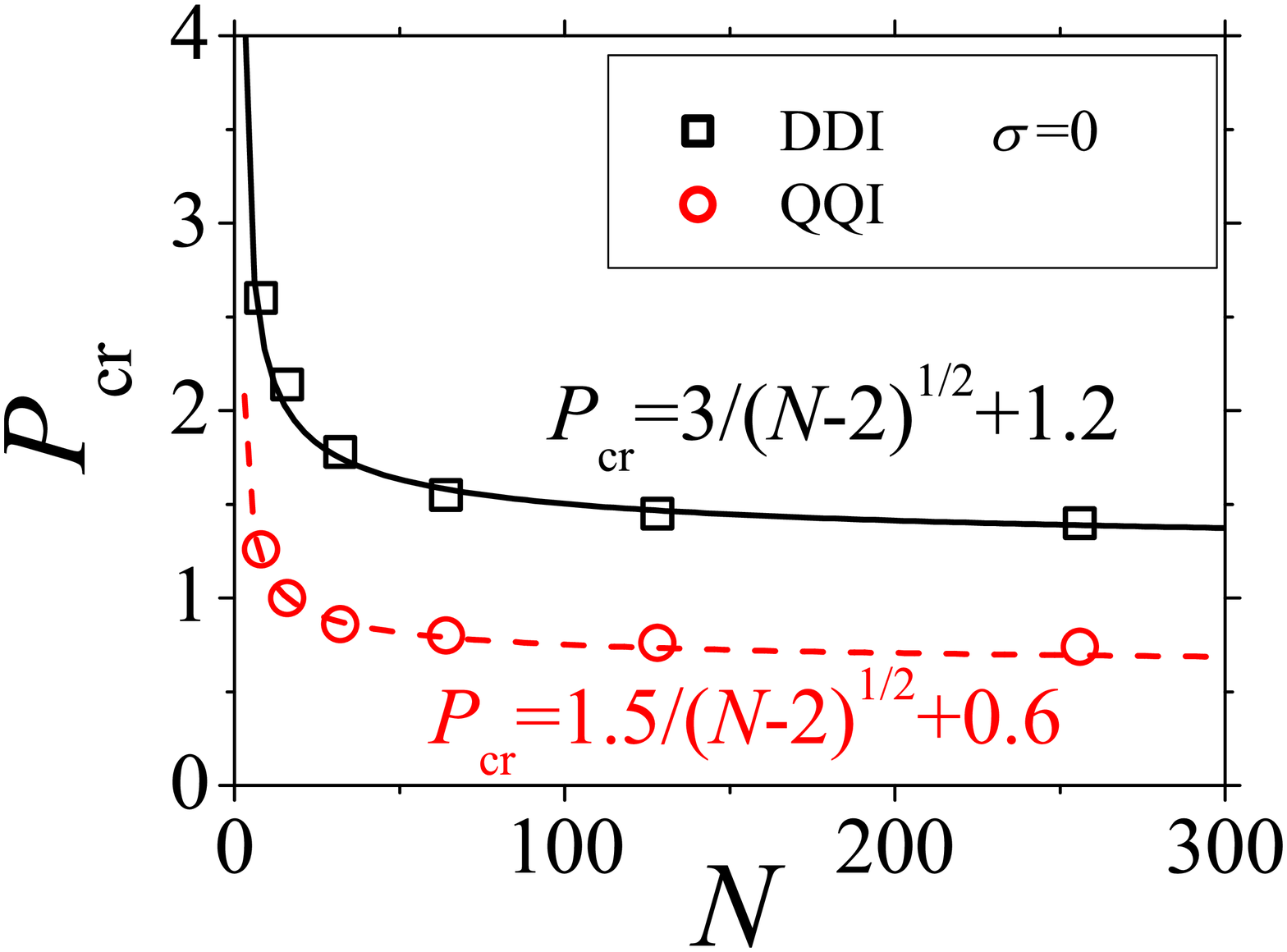}}
\subfigure[]{\label{fig3c}
\includegraphics[scale=0.22]{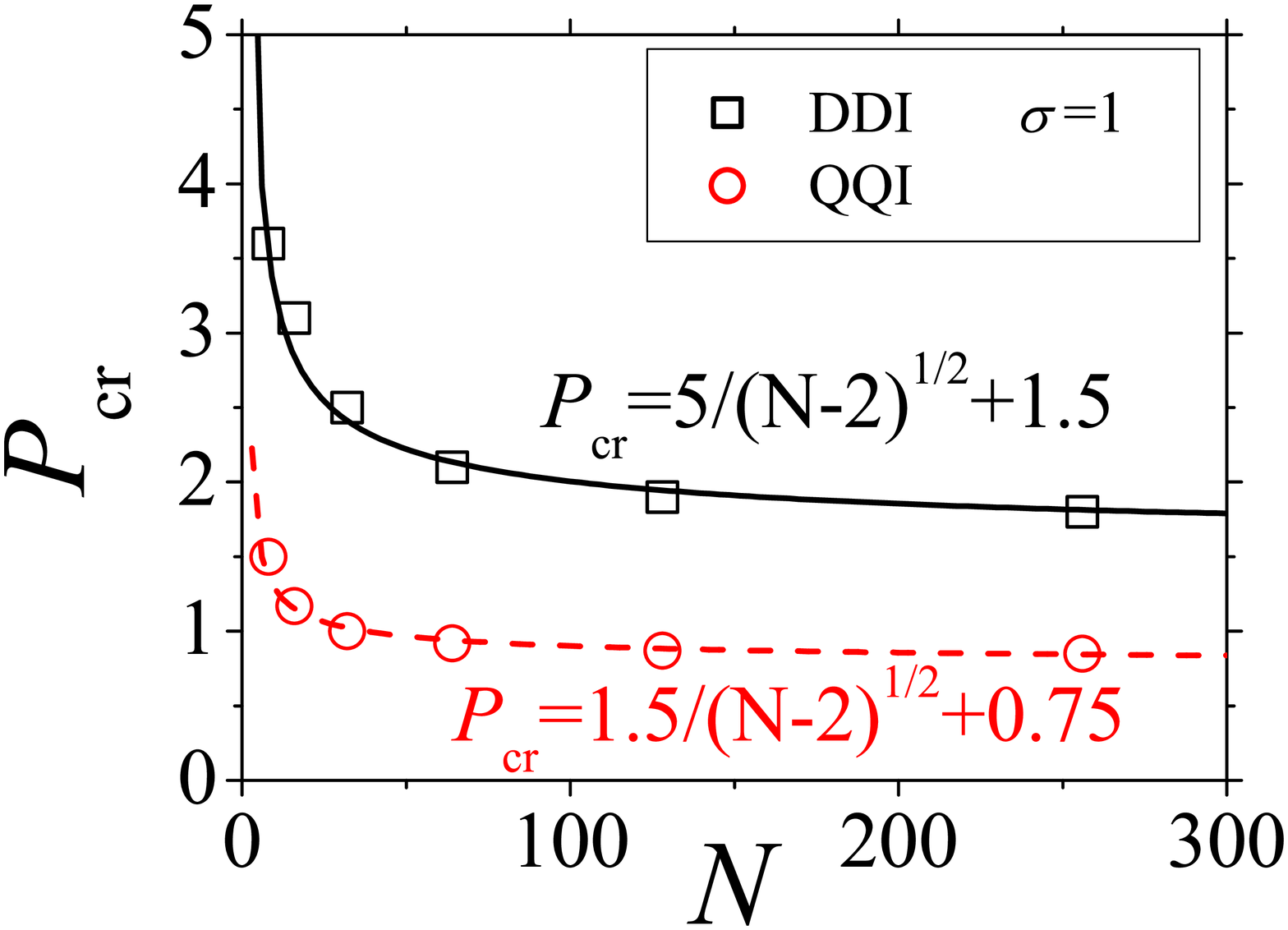}}
\subfigure[]{\label{fig3d}
\includegraphics[scale=0.22]{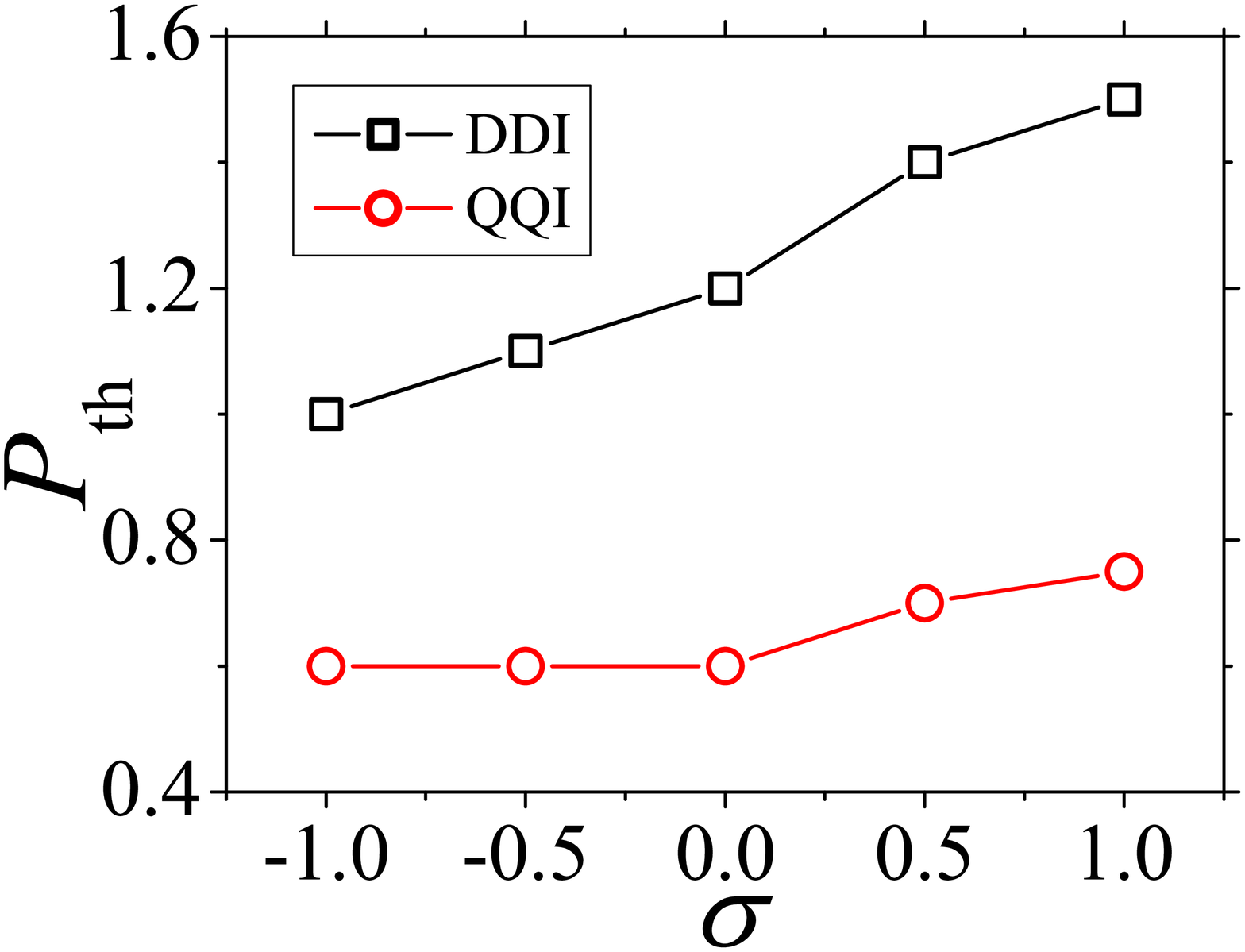}}
\caption{(Color online) The minimum norm necessary for the formation of 2D
discrete solitons, $P_{\mathrm{cr}}$, under the action of the dipole-dipole
(DDI) or quadrupole-quadrupole (QQI) interactions, as a function of the size
of the solution domain $N$, for (a) $\protect\sigma =-1$, the attractive
contact nonlinearity; (b) $\protect\sigma =0$, zero contact nonlinearity;
(c) $\protect\sigma =1$, repulsive contact nonlinearity. Formulas written in
the panels represent empiric fits of the plots to simple analytical
expressions. (d) The threshold value of the norm, which corresponds to $P_{%
\mathrm{cr}}$ at $N\rightarrow \infty $, versus the strength of the contact
nonlinearity. }
\label{thresholdnorm}
\end{figure}

As shown in Fig. \ref{fig3a}-\ref{fig3c} for different fixed values of
strength $\sigma $ of the contact nonlinearity in Eq. (\ref{discr}), $P_{%
\mathrm{cr}}$ depends on the size of the solution domain, $N$, because in
the case of relatively small $N$ only strongly self-trapped solitons, with
sufficiently large $P$, may be narrow enough to fit into\ the domain. The
\textit{threshold} for the formation of the 2D solitons, i.e., the absolute
minimum of $P_{\mathrm{cr}}$, can be identified as $P_{\mathrm{th}}=P_{%
\mathrm{cr}}(N\rightarrow \infty $). Fitting formulas displayed in panels %
\ref{fig3a}-\ref{fig3c} help to identify the respective values of $P_{%
\mathrm{th}}$.

The overall dependence of $P_{\mathrm{th}}$ on $\sigma $ is displayed in
Fig. \ref{fig3d}. The increase of $P_{\mathrm{th}}$ with $\sigma $ is
naturally explained by the competition between the long-range
self-attraction and contact self-repulsion at $\sigma >0$ (or insufficient
self-attraction at $\sigma <0$).

The results summarized in Fig. \ref{thresholdnorm} demonstrate that the QQI
offers an advantage, in comparison with the DDI, for the formation of
solitons, as the respective scaled formation threshold is considerably lower
(roughly, by a factor of $2$) and, in addition to that, in the case of QQI\
the threshold is weakly affected by the competition with the contact
nonlinearity, unlike the DDI model, where it is an important issue \cite%
{Pfau,Pfau2}.

\begin{figure}[tbp]
\centering\subfigure[] {\label{fig4a}
\includegraphics[scale=0.18]{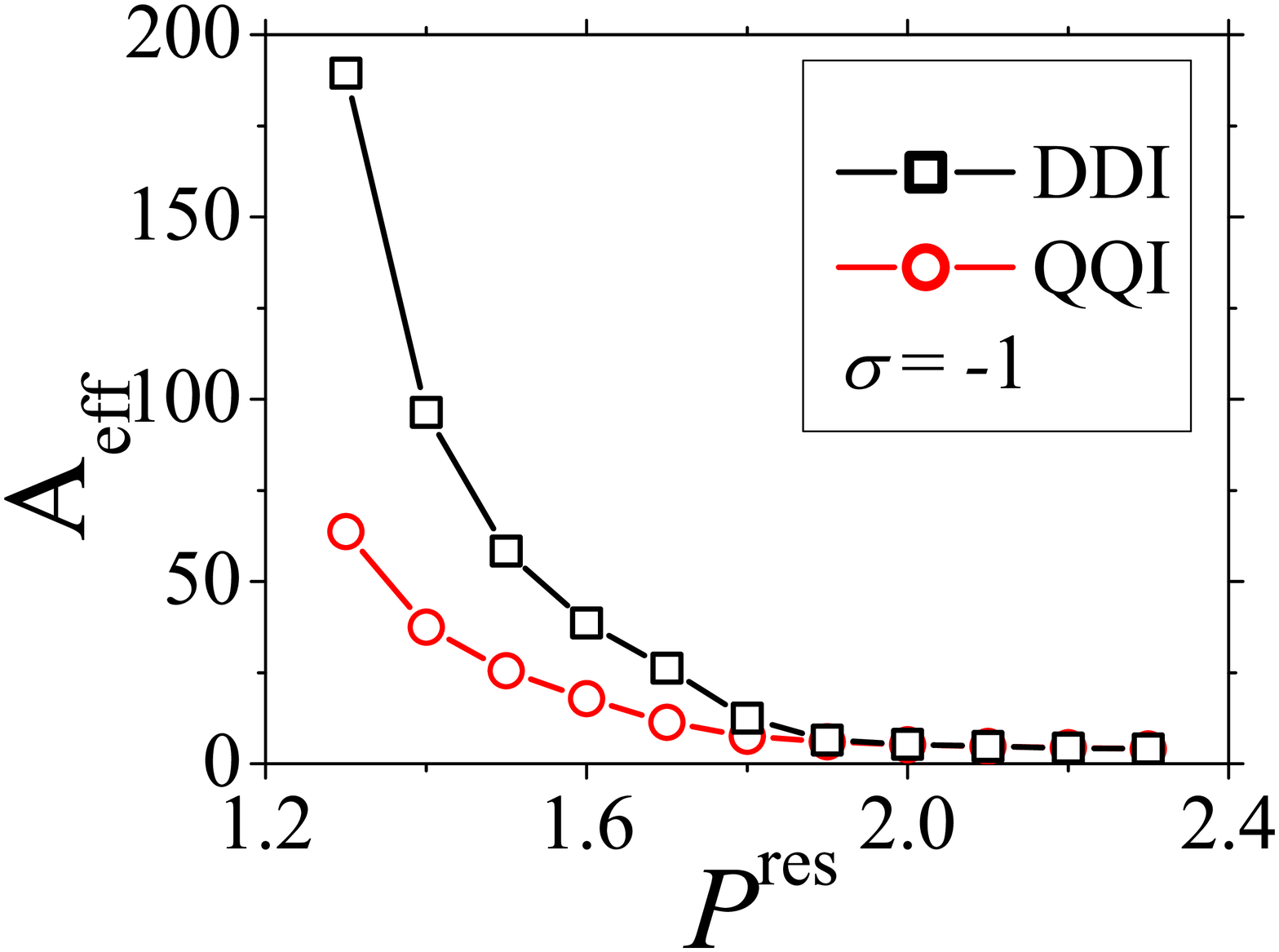}}%
\subfigure[] {\label{fig4b}
\includegraphics[scale=0.18]{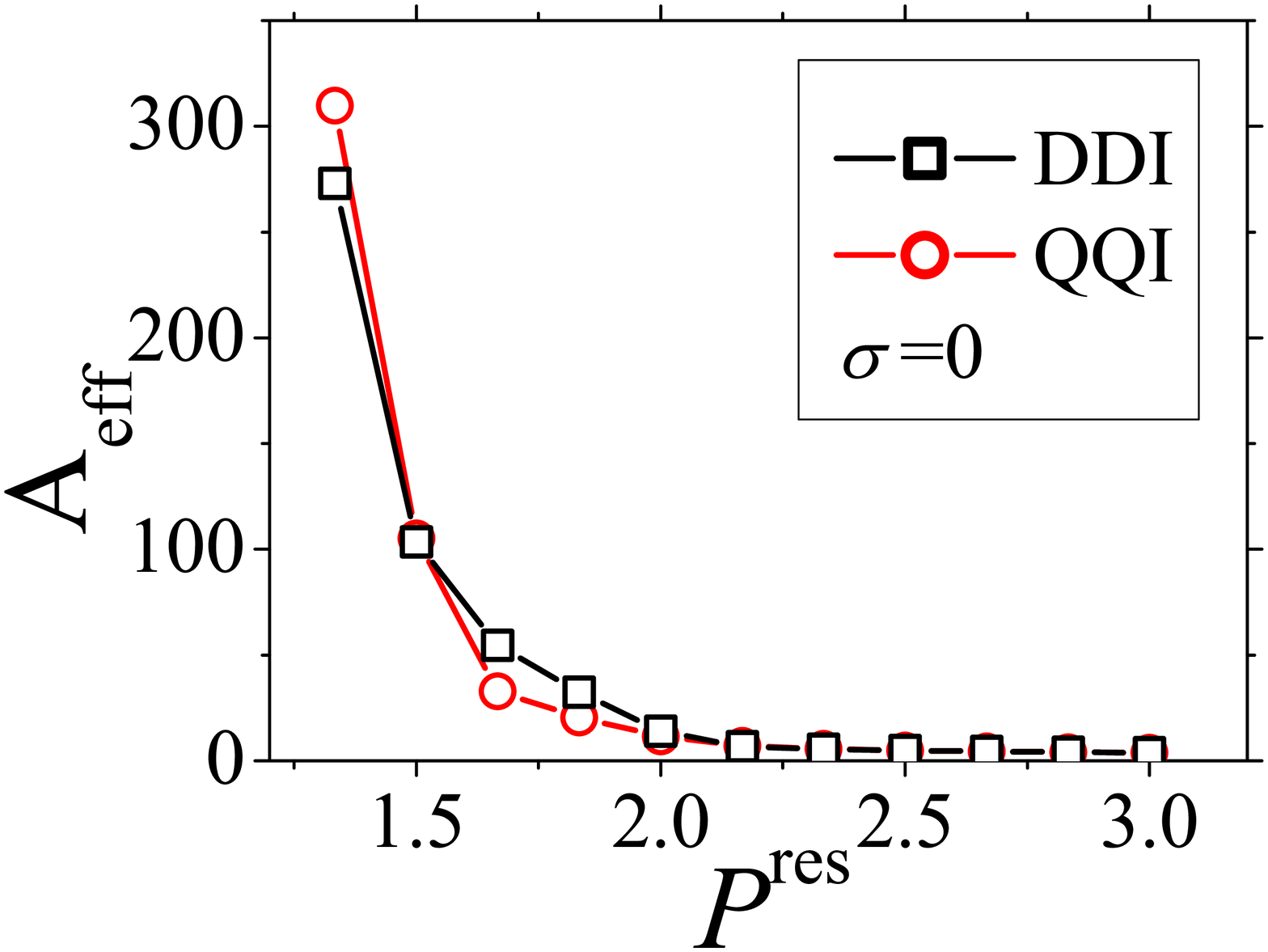}}
\subfigure[]{\label{fig4c}
\includegraphics[scale=0.18]{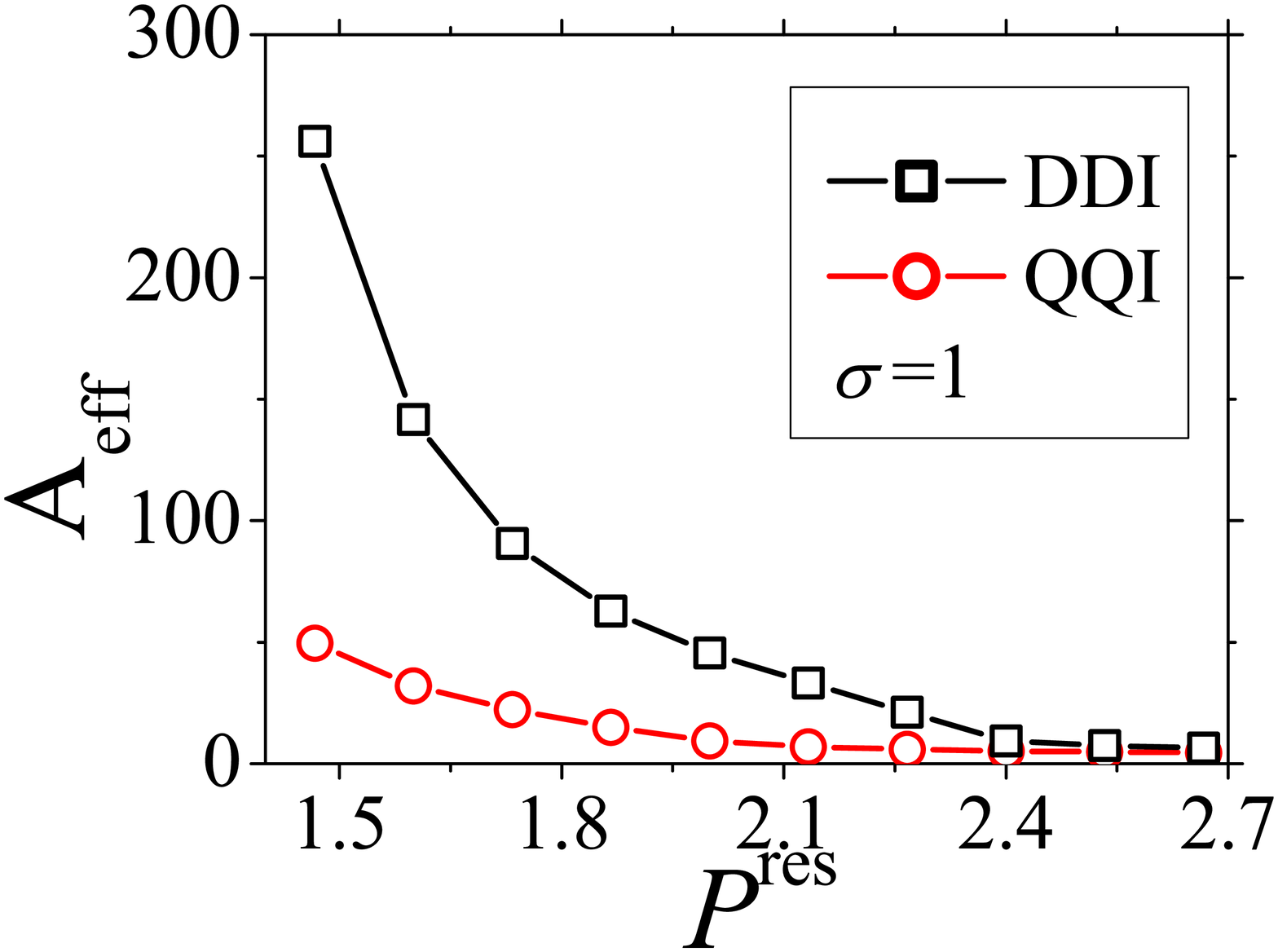}}
\caption{(Color online) Effective area $\mathrm{A_{eff}}$ [see Eq. (\protect
\ref{Aeff})] versus the rescaled norm [see Eq. (\protect\ref{rescal})] for
the solitons supported by the QQI and DDI in the combination with $\protect%
\sigma =-1$ (a), $\protect\sigma =0$ (b), and $\protect\sigma =1$ (c), i.e.,
the attractive, zero, or repulsive contact interactions. The numerical
domain here is $64\times 64$.}
\label{Comaeff}
\end{figure}

\begin{figure}[tbp]
\centering\subfigure[] {\label{fig5a}
\includegraphics[scale=0.18]{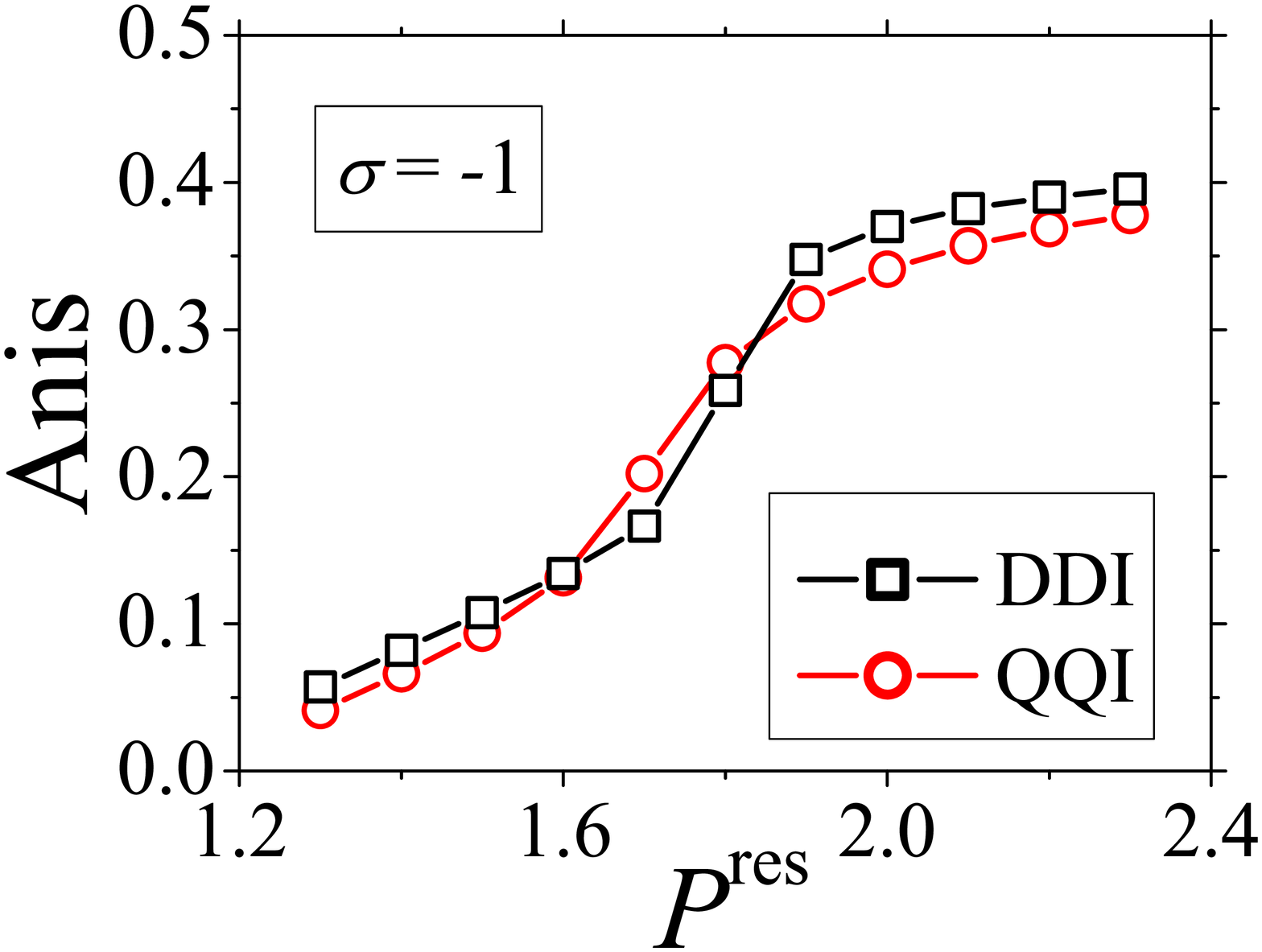}}%
\subfigure[] {\label{fig5b}
\includegraphics[scale=0.18]{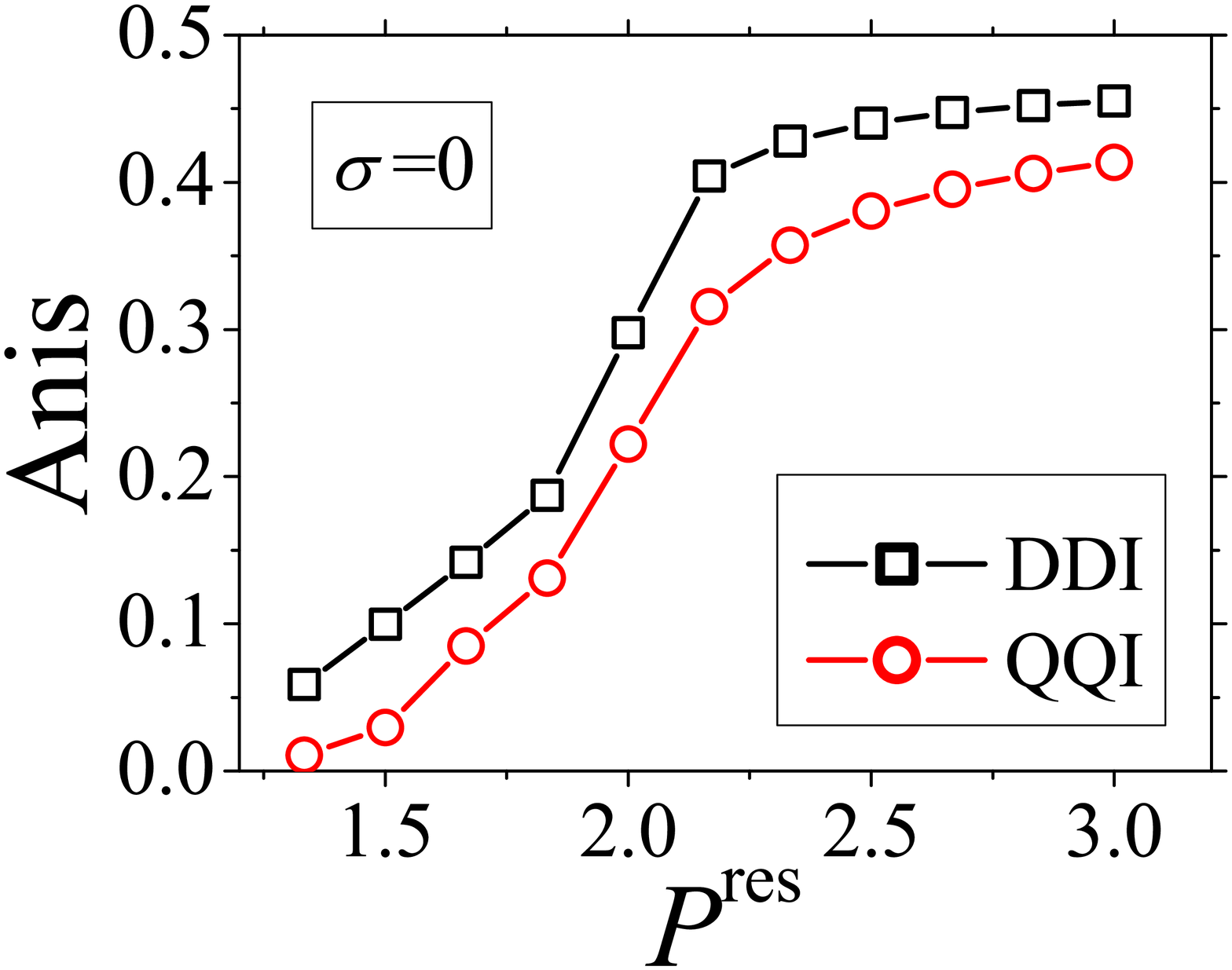}}
\subfigure[]{\label{fig5c}
\includegraphics[scale=0.18]{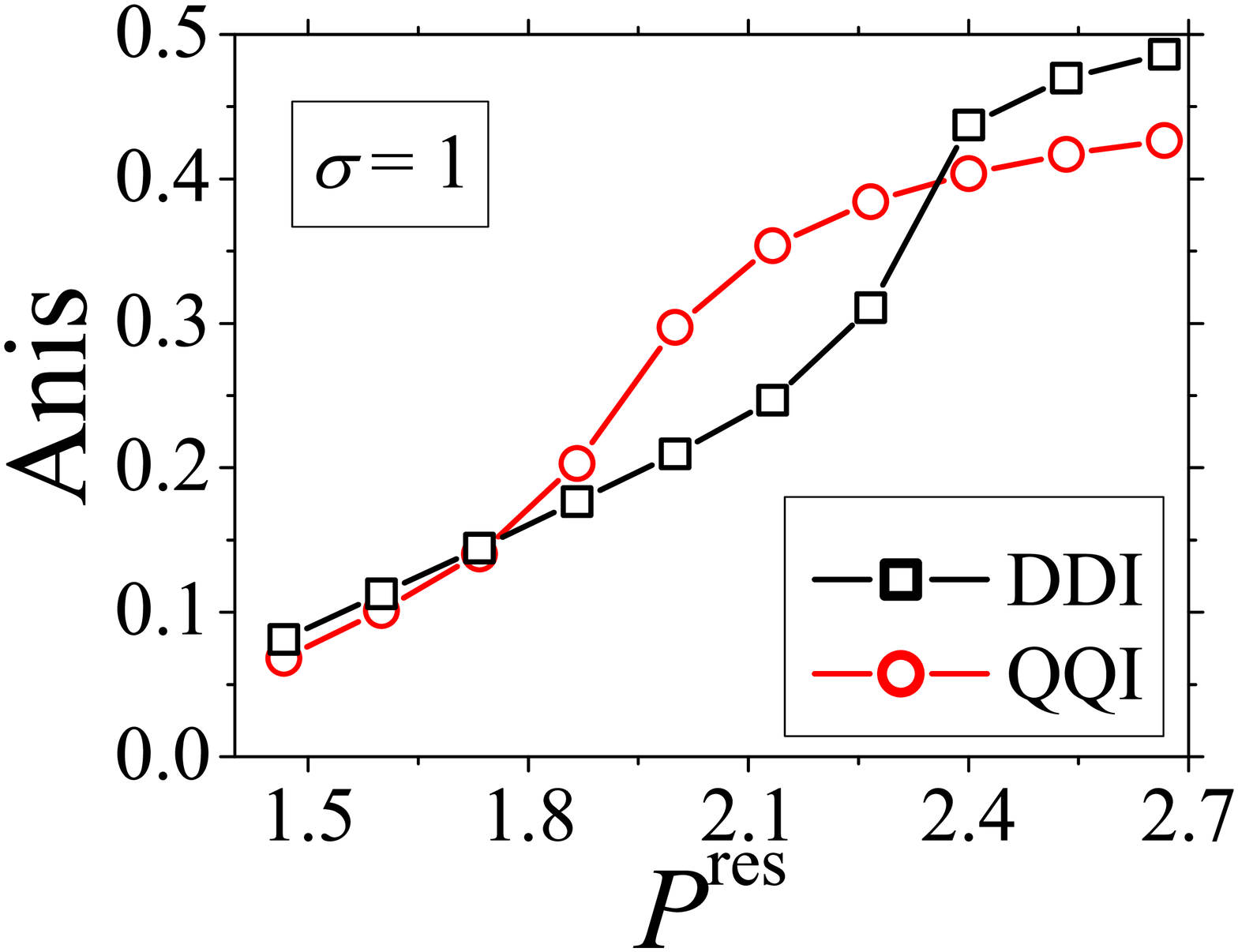}}
\caption{(Color online) The same as in Fig. \protect\ref{Comaeff}, but for
anisotropy $\mathrm{Anis}$, see Eq. (\protect\ref{Anis}). }
\label{ComAnis}
\end{figure}

To further quantify the solitons, we define their effective area, $\mathrm{%
A_{eff}}$, and structural anisotropy between the horizontal
(``hor") and vertical (``vert") directions, $%
\mathrm{Anis}$, as follows:
\begin{equation}
\mathrm{A_{eff}}={\frac{\left( \sum_{m,n}|\phi _{mn}|^{2}\right) ^{2}}{%
\sum_{m,n}|\phi _{mn}|^{4}}};  \label{Aeff}
\end{equation}%
\begin{gather}
\mathrm{Anis}={\frac{\left\vert D_{\mathrm{hor}}-D_{\mathrm{vert}%
}\right\vert }{\left\vert D_{\mathrm{hor}}+D_{\mathrm{vert}}\right\vert }},~
\label{Anis} \\
D_{\mathrm{hor}}\equiv {\frac{\left( \sum_{m}|\phi _{m,0}|^{2}\right) ^{2}}{%
\sum_{m}|\phi _{m,0}|^{4}}},~D_{\mathrm{vert}}\equiv {\frac{\left(
\sum_{n}|\phi _{0,n}|^{2}\right) ^{2}}{\sum_{n}|\phi _{0,n}|^{4}}},\quad
\notag
\end{gather}%
where it is assumed that the soliton's center is fixed at site $m=n=0$. In
Figs. \ref{Comaeff} and \ref{ComAnis}, $\mathrm{A_{eff}}$ and $\mathrm{Anis}$
are plotted, along with the solitons' chemical potential, versus the \textit{%
rescaled norm}, which is defined as
\begin{equation}
P^{\mathrm{res}}\equiv P/P_{\mathrm{th}}^{\mathrm{(QQ,DD)}},  \label{rescal}
\end{equation}%
where $P_{\mathrm{th}}^{\mathrm{(QQ,DD)}}$ is taken, severally, from Fig. %
\ref{fig3d} as as the respective threshold value for the QQI and DDI models.
\begin{figure}[tbp]
\centering\subfigure[] {\label{fig6a}
\includegraphics[scale=0.18]{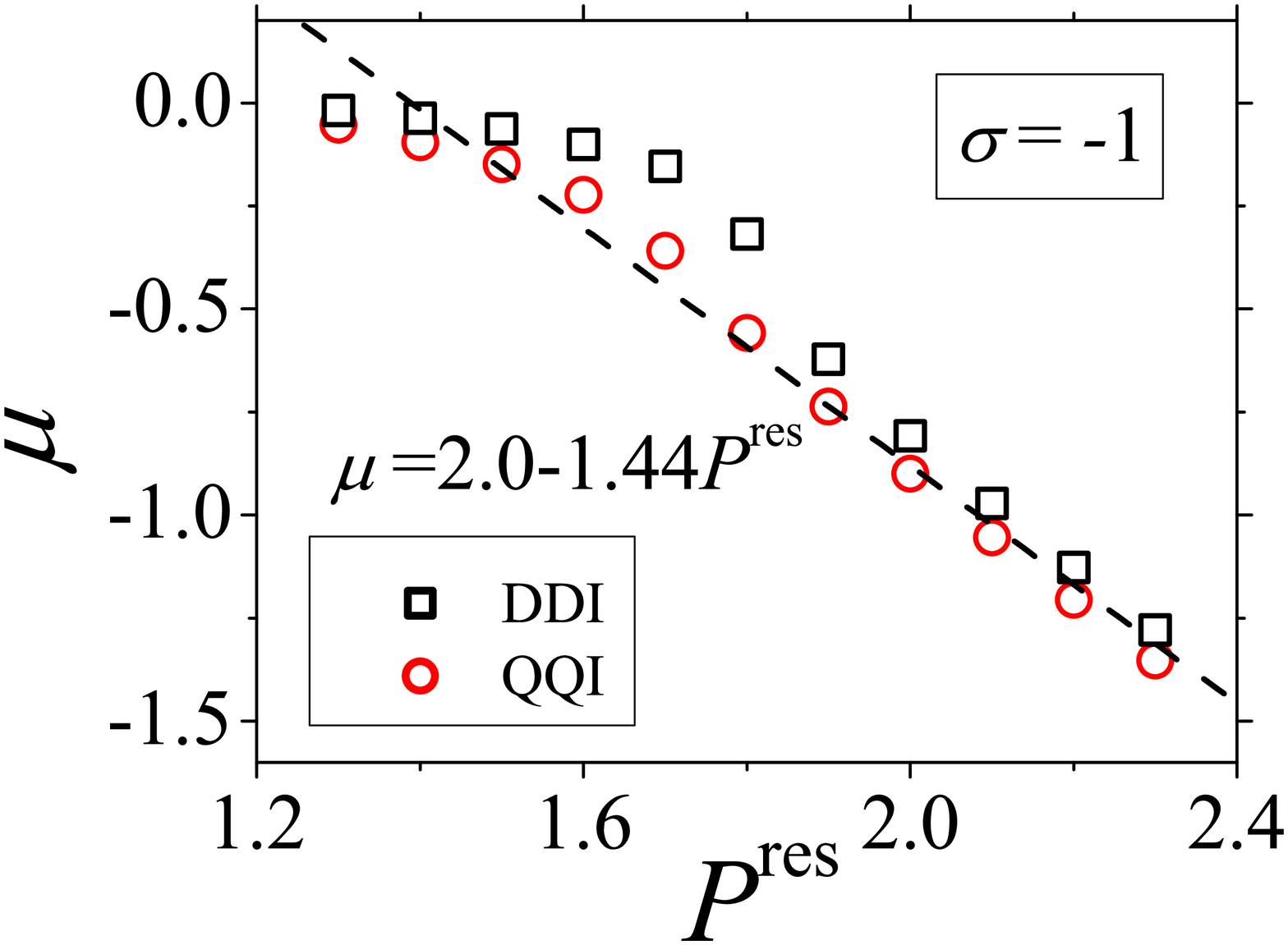}}%
\subfigure[] {\label{fig6b}
\includegraphics[scale=0.18]{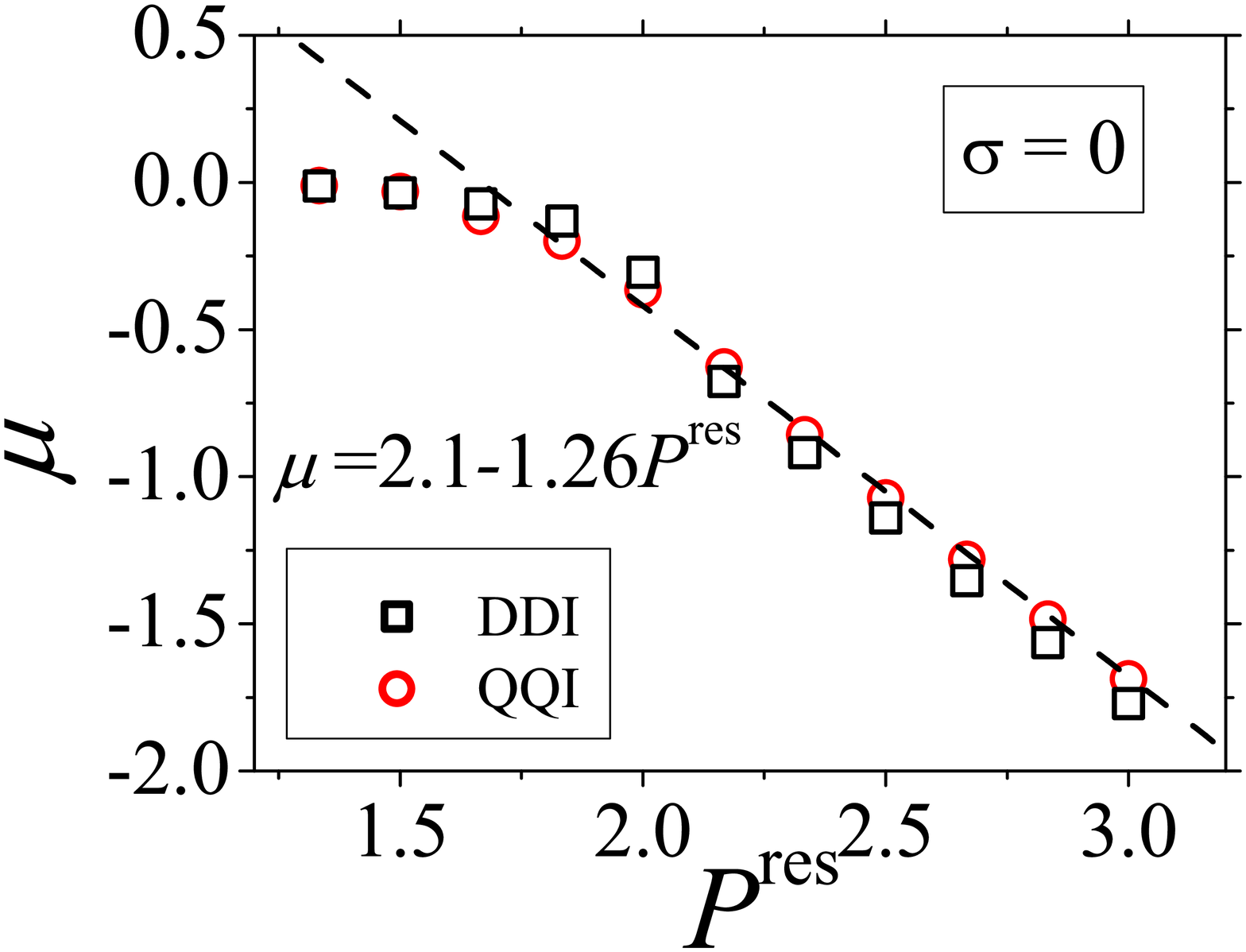}}
\subfigure[]{\label{fig6c}
\includegraphics[scale=0.18]{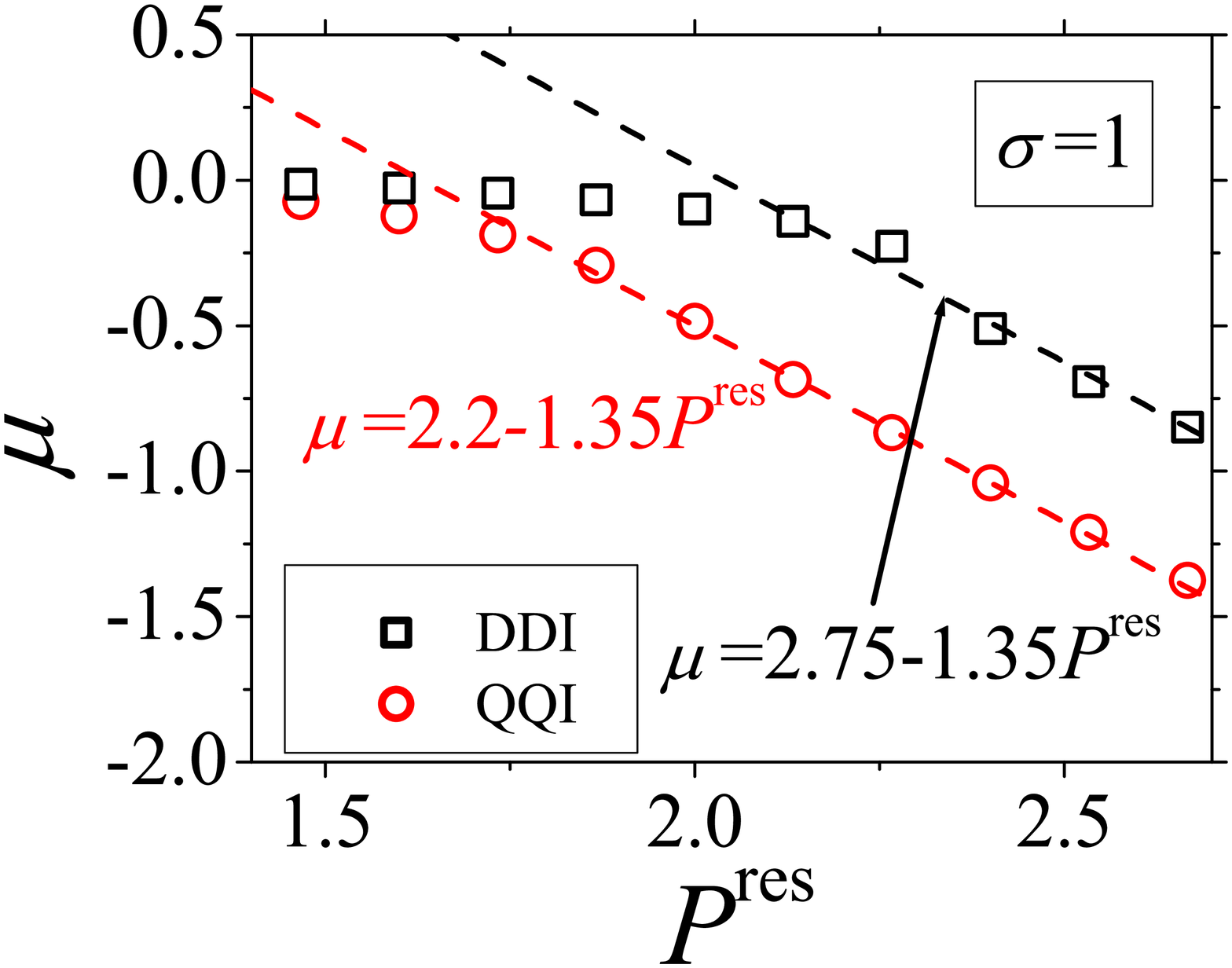}}
\caption{(Color online) The same as in Figs. \protect\ref{Comaeff} and
\protect\ref{ComAnis}, but for chemical potential $\protect\mu $, see Eq. (%
\protect\ref{mu}). The formulas present\ analytical fits for quasi-linear
portions of the $\protect\mu (P)$ dependences.}
\label{Commu}
\end{figure}

Figures \ref{Comaeff} and \ref{ComAnis} demonstrate that the increase of the
norm makes the solitons more tightly localized and more anisotropic, hence
the long-range interactions, QQI or DDI, become dominant over the isotropic
local interaction, for ``heavier" solitons. Further, it is worthy to note
that $\mu (P)$ dependences in Fig. \ref{Commu} satisfy the
Vakhitov-Kolokolov criterion, $d\mu /dP<0$, which is a well-known necessary
stability condition for modes supported by self-attractive nonlinearities
\cite{VK0,VK1}. In particular, this is true in the case of $\sigma =+1 $,
when the attractive long-range interactions clearly beat the contact
self-repulsion.

\section{Mobility and collisions of the discrete solitons}

\subsection{Setting the solitons in motion}

Mobility of discrete solitons is a basic issue in the theory of nonlinear
dynamical lattices \cite{Kevrekid}. In 2D lattices with the onsite (contact)
nonlinearity, the mobility crucially depends of the type of the
nonlinearity. The cubic onsite term, which corresponds to the critical
collapse in the 2D continuum limit \cite{VK1}, gives rise to immobile
discrete solitons strongly pinned to the underlying lattice. On the other
hand, subcritical nonlinear terms, such as saturable \cite{Johansson} or
quadratic \cite{Hadi}, can readily create mobile 2D lattice solitons. In the
1D setting, it has been demonstrated that the DDI helps to enhance the
mobility of discrete solitons \cite{Belgrade,Zhihuan}. The mobility and its
consequences, such as collisions, were not studied previously on 2D lattices
with long-range intersite interactions.

Here we focus on the mobility of 2D discrete solitons built solely by the
long-range interactions (QQI), setting $\sigma =0$ in Eq. (\ref{discr}).
Because the system is anisotropic, we separately consider the initiation of
the soliton motion by kicks applied in the horizontal and vertical
directions:
\begin{equation}
\psi _{mn}^{\mathrm{(hor)}}(t=0)=\phi _{mn}e^{i\eta m},~\psi _{mn}^{\mathrm{%
(vert)}}(t=0)=\phi _{mn}e^{i\eta n},  \label{kick}
\end{equation}%
where $\eta $ is the strength of the kick and $\phi _{mn}$ is the stationary
soliton solution [recall it is prolate in the horizontal ($m$) direction,
see Fig. \ref{examplessolution}].

\begin{figure}[tbp]
\centering\subfigure[] {\label{fig7a}
\includegraphics[scale=0.18]{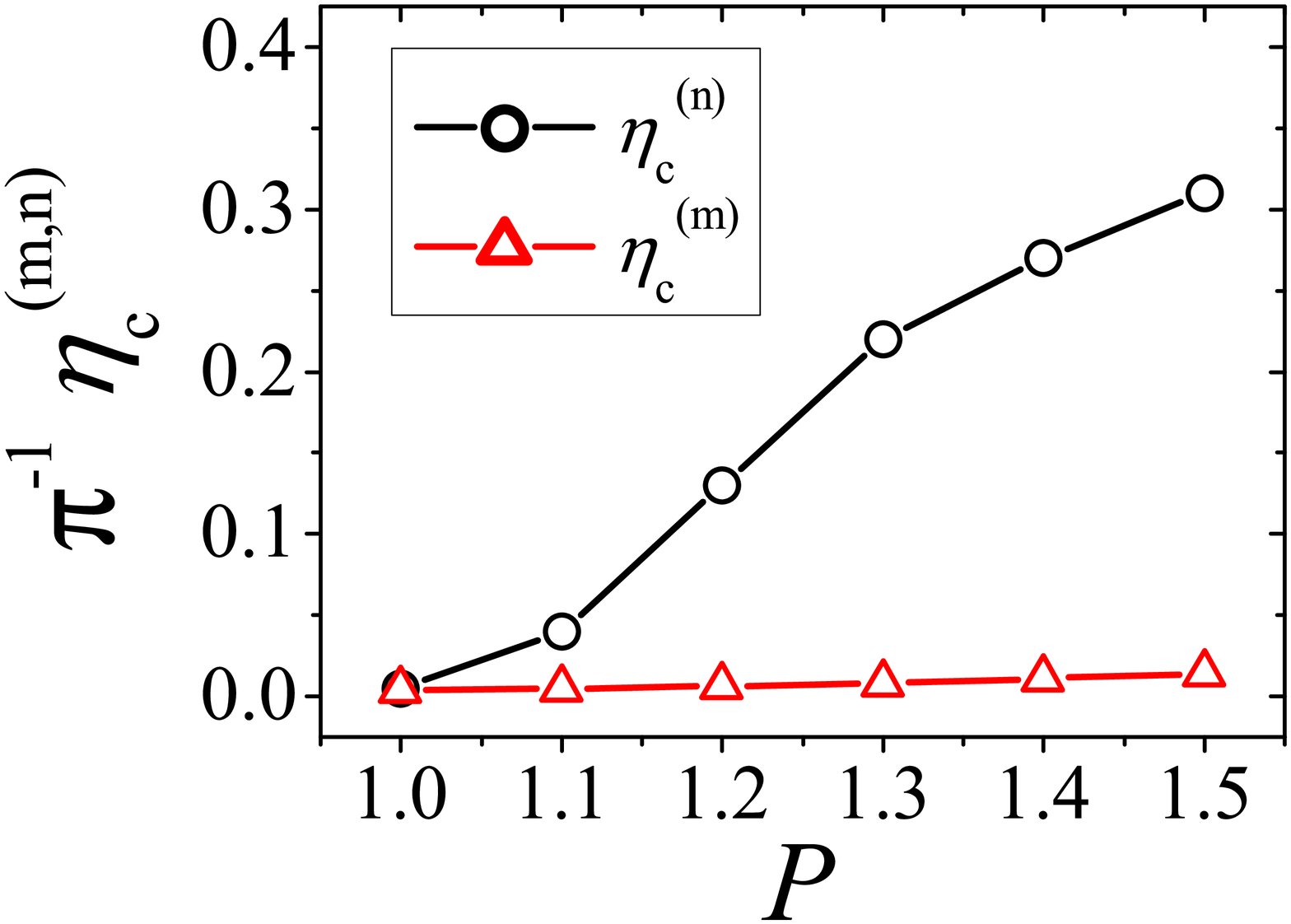}}%
\subfigure[] {\label{fig7b}
\includegraphics[scale=0.18]{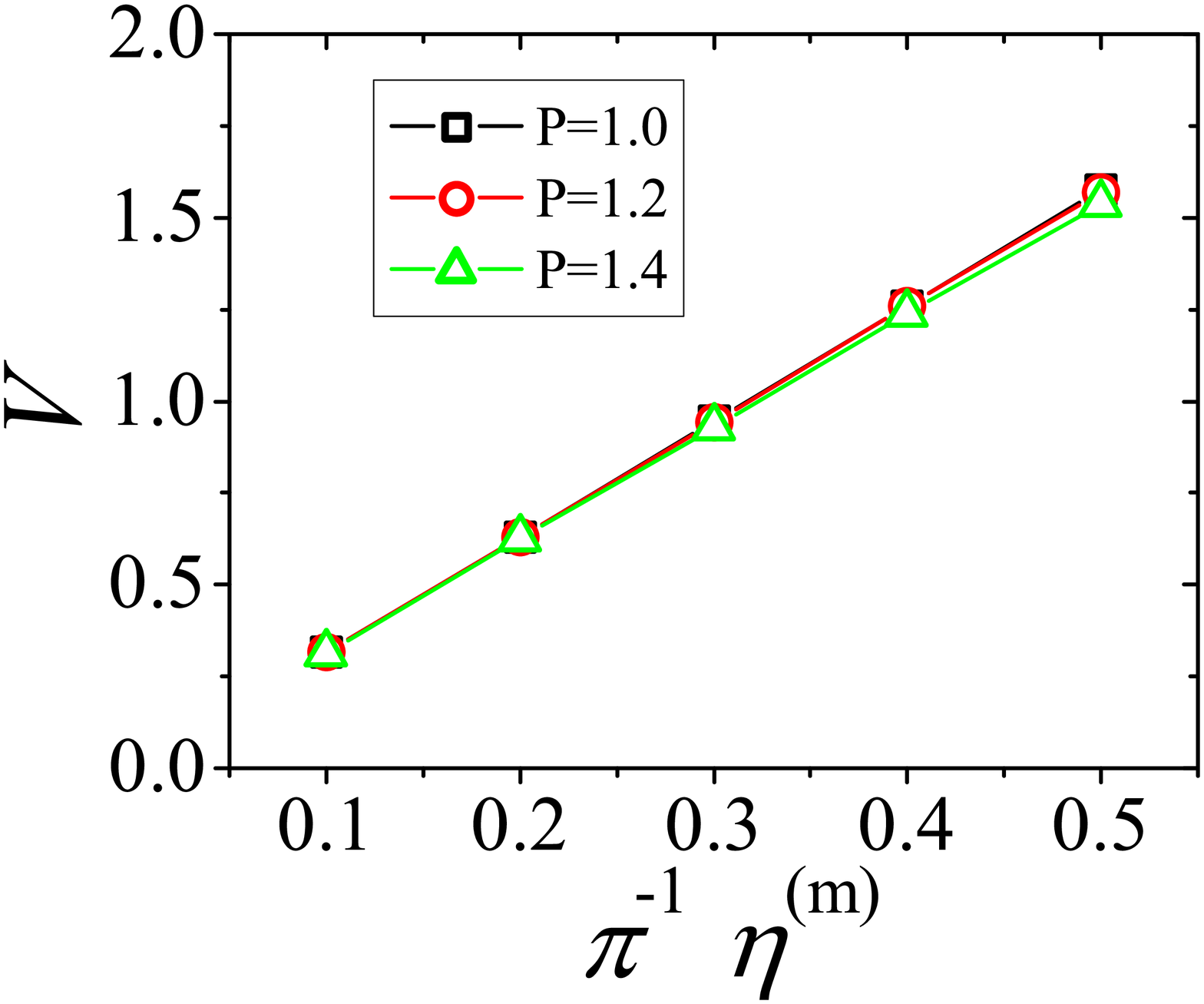}}
\subfigure[]{\label{fig7c}
\includegraphics[scale=0.18]{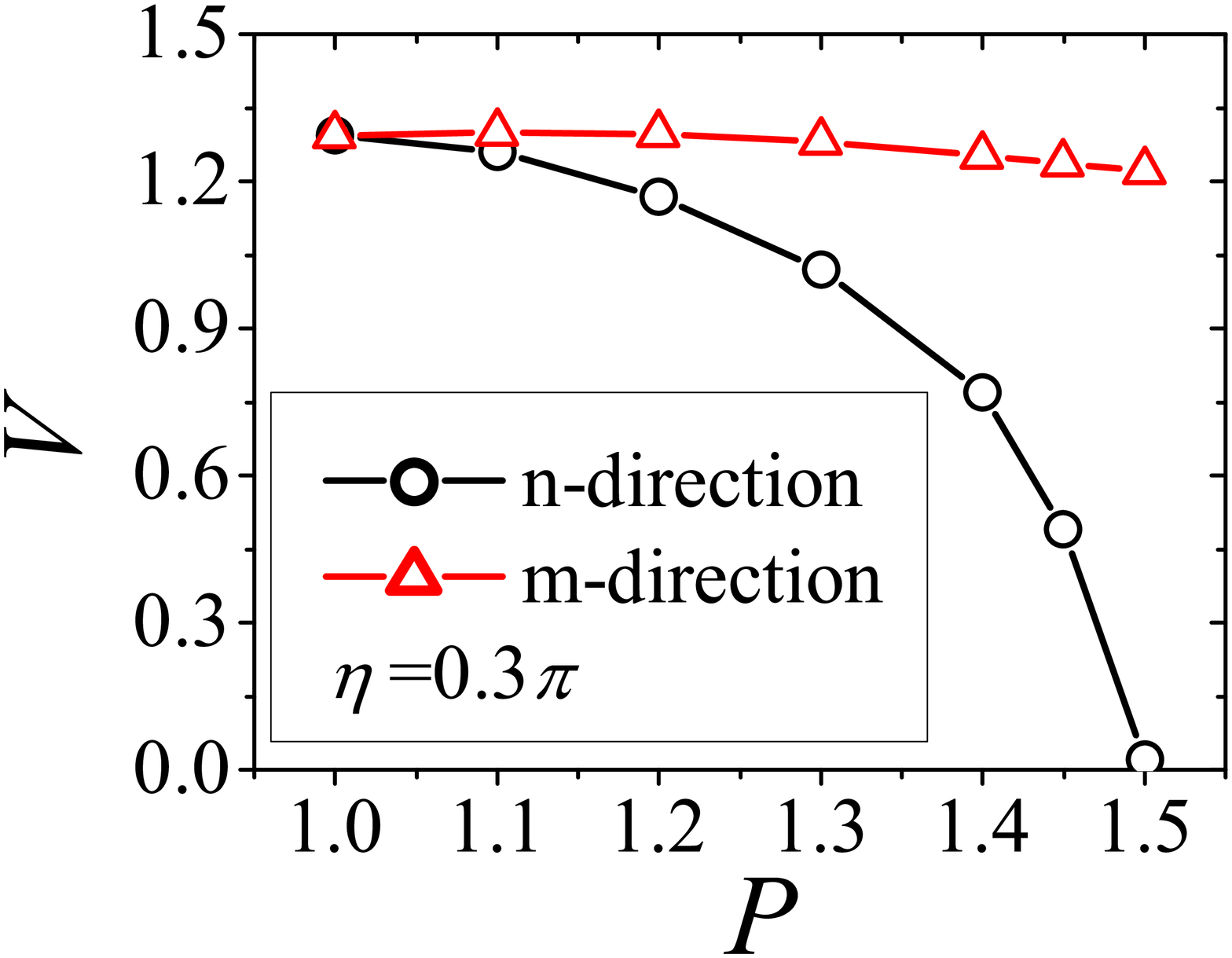}}
\caption{(Color online) (a) The minimum size of the kick, necessary to set a
soliton in motion in the horizontal and vertical directions [$\protect\eta _{%
\mathrm{c}}^{(m)}$ and $\protect\eta _{\mathrm{c}}^{(n)}$, respectively], as
a function of the soliton's norm, $P$. (b) Velocity of the soliton moving in
the horizontal direction versus the strength of the kick (at $\protect\eta >%
\protect\eta _{\mathrm{c}}^{(m)}$) for different fixed values of the norm.
(c) Velocity of the soliton moving in the horizontal ($m$) and vertical ($n$%
) directions versus $P$ for a fixed kick's strength, $\protect\eta =0.3>%
\protect\eta _{\mathrm{c}}^{(m,n)}$.}
\label{kickmoveproperty}
\end{figure}

Simulations demonstrate that, to set the solitons in the state of persistent
motion, the kick must exceed a finite threshold value, $\eta _{\mathrm{c}}$.
The kick with $\eta <\eta _{\mathrm{c}}$ may only shift the soliton from the
initial position by a finite distance (i.e., the soliton starts to move but
then comes to a halt). Figure \ref{fig7a} shows the threshold as a function
of of the soliton's norm, $P$. It is seen that the dependence is strongly
anisotropic: the horizontal threshold, $\eta _{\mathrm{c}}^{(m)}$, is very
small and almost does not depend on $P$, while its vertical counterpart, $%
\eta _{\mathrm{c}}^{(n)}$, is much larger, and grows roughly linearly with $%
P $.

\begin{figure}[tbp]
\centering\subfigure[] {\label{fig8a}
\includegraphics[scale=0.35]{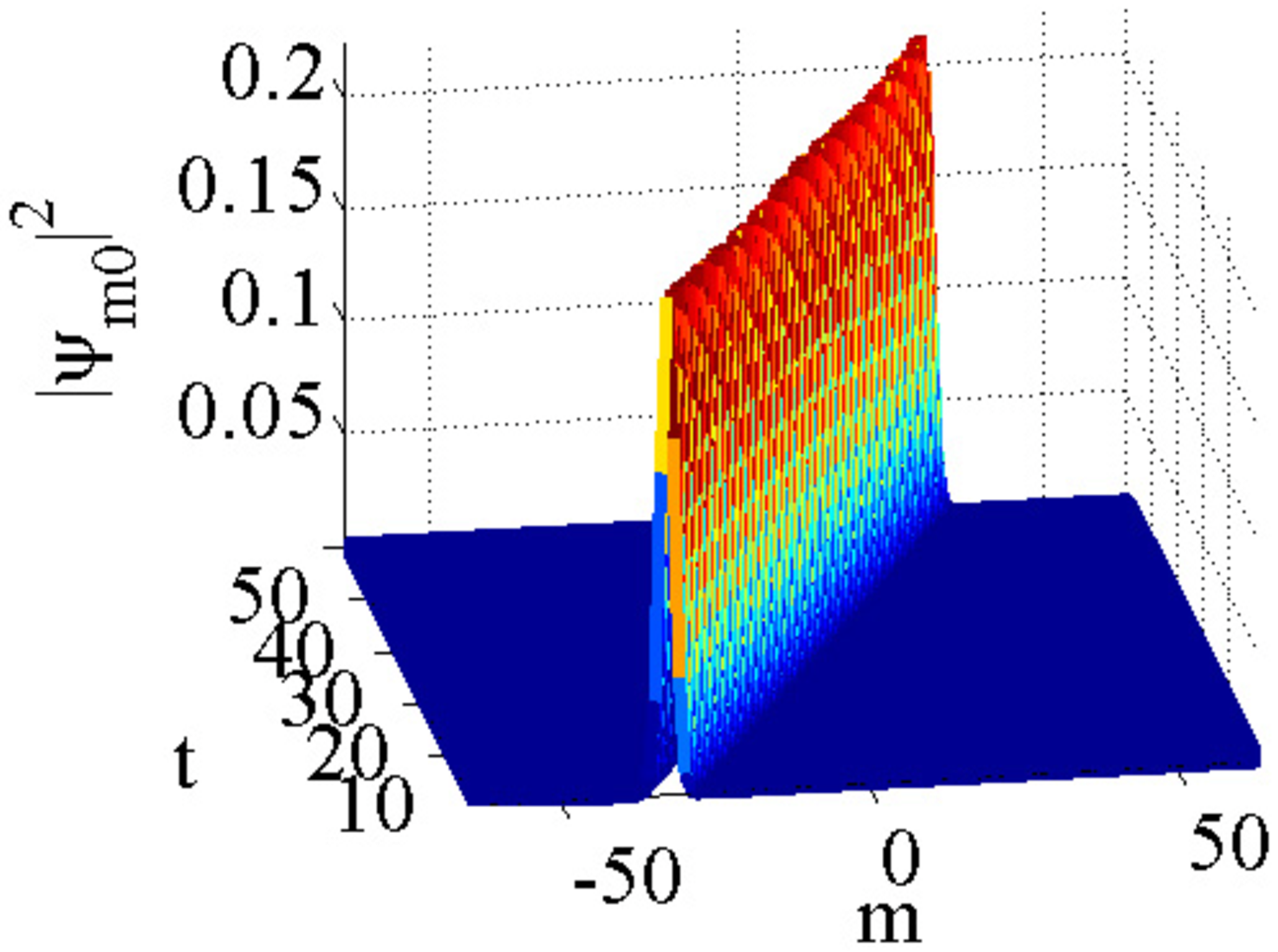}}%
\subfigure[] {\label{fig8b}
\includegraphics[scale=0.35]{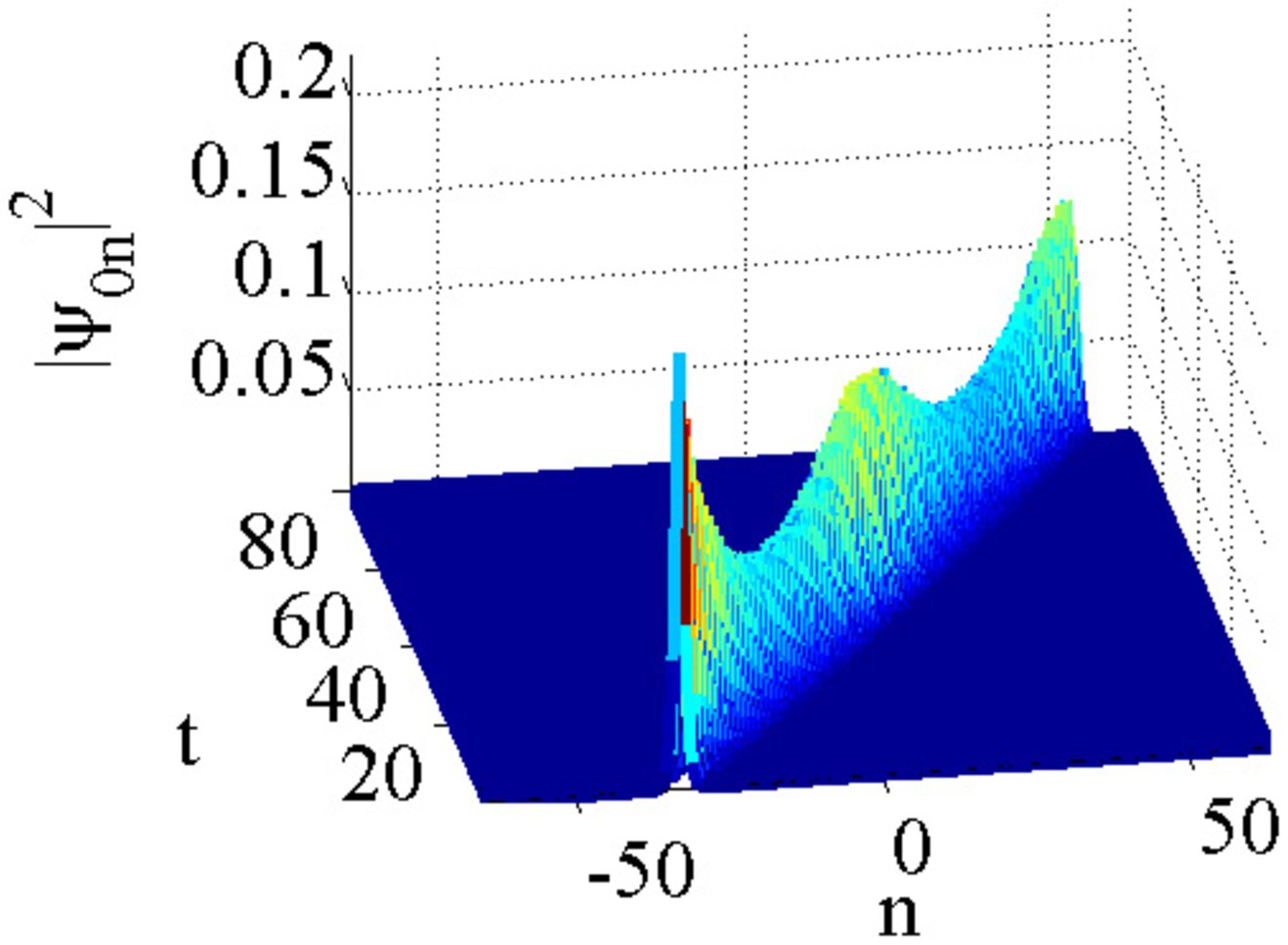}}
\subfigure[]{\label{fig8c}
\includegraphics[scale=0.32]{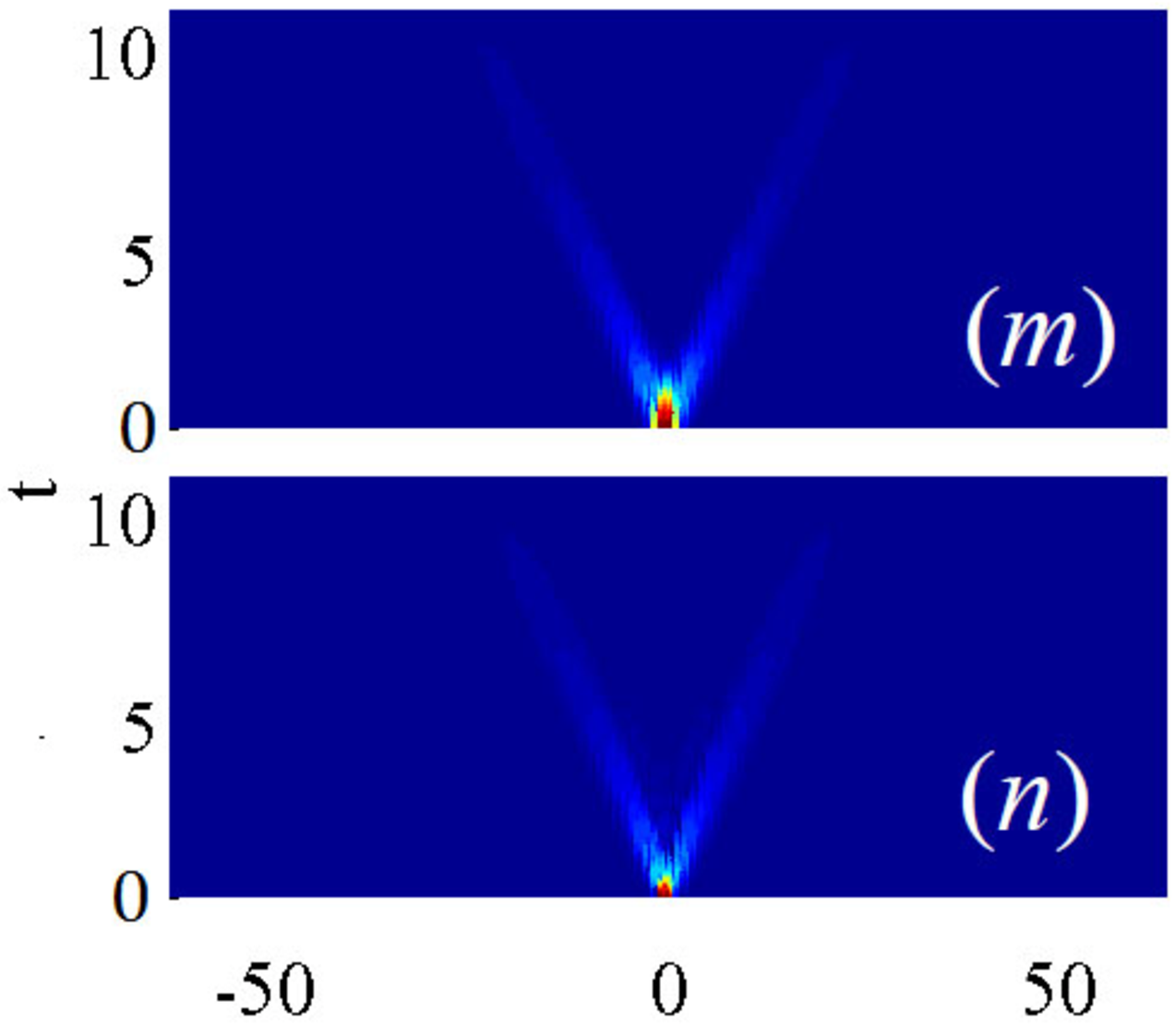}}
\caption{(Color online) The evolution of a soliton with norm $P=1.2$, kicked
with strength $\protect\eta =0.4\protect\pi $ in the $m$- (horizontal)
direction (a) and $n$- (vertical) direction (b). The evolution is displayed
in cross sections $(n=0)$ and $(m=0)$, respectively. (c) Destruction of the
soliton by the staggering kick, $\protect\eta =\protect\pi $, applied in
either direction.}
\label{Movingexample1}
\end{figure}

Obviously, the size of the kick in Eq. (\ref{kick}) is limited to $\eta \leq
\pi $. In the limit case of $\eta =\pi $, the kick makes the soliton \textit{%
staggered}, rather than moving, which results in destruction of the kicked
soliton. Typical examples of the evolution of the kicked solitons are
displayed in Fig. \ref{Movingexample1}. In panel \ref{fig8a}, the soliton
moving in the horizontal direction is robust, while in panel \ref{fig8b} the
same soliton, kicked with the same $\eta $ but vertically, is moving in a
strongly perturbed oscillatory state. The trend to the destruction of the
kicked soliton becomes apparent at $\eta >\pi /2$. The destruction in the
case of $\eta =\pi $ is shown in panel \ref{fig8c}.

Dependences of the velocity of the moving soliton on the kick's strength ($%
\eta $) and direction ($m$ or $n$ -- horizontal or vertical), and on the
soliton's power ($P$), are presented in Figs. \ref{fig7b} and \ref{fig7c}.
The former figure shows that, similar to the usual properties of
nonlinear-Schr\"{o}dinger\ solitons in the continuous medium, the velocity
of the horizontally kicked soliton increases as a linearly function of $\eta
$ (with an offset at $V=0$ corresponding to the threshold value, $\eta _{%
\mathrm{c}}^{(m)}$), and practically does not depend on the norm. On the
other hand, Fig. \ref{fig7c} demonstrates that velocity of the established
motion in the vertical direction strongly depends on the soliton's norm, the
heavier solitons being much less motile. Thus, the mobility of the 2D
solitons maintained by the QQI is strongly anisotropic.

\begin{figure}[tbp]
\centering\subfigure[] {\label{fig9a}
\includegraphics[scale=0.22]{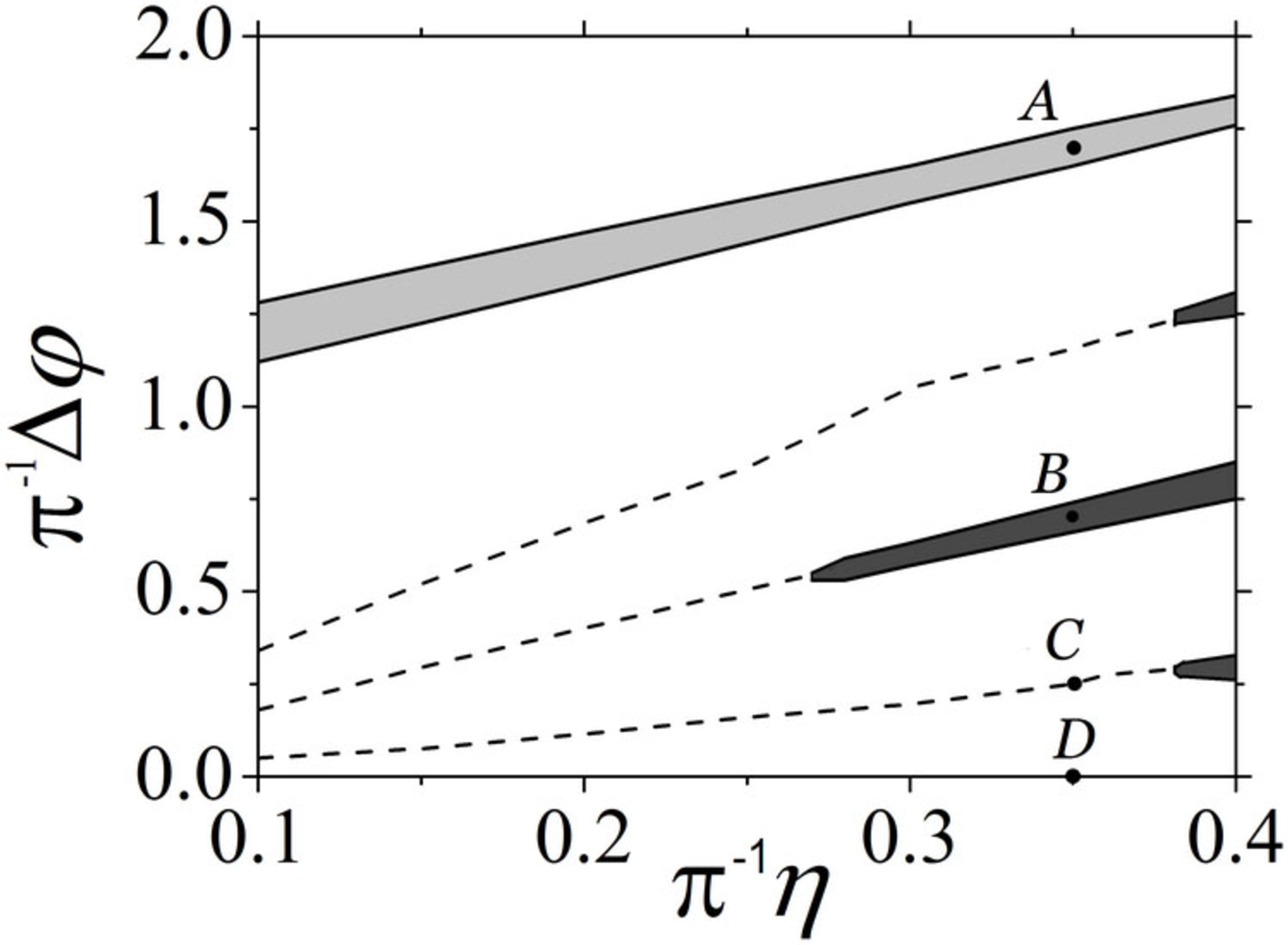}}
\subfigure[] {\label{fig9b}
\includegraphics[scale=0.22]{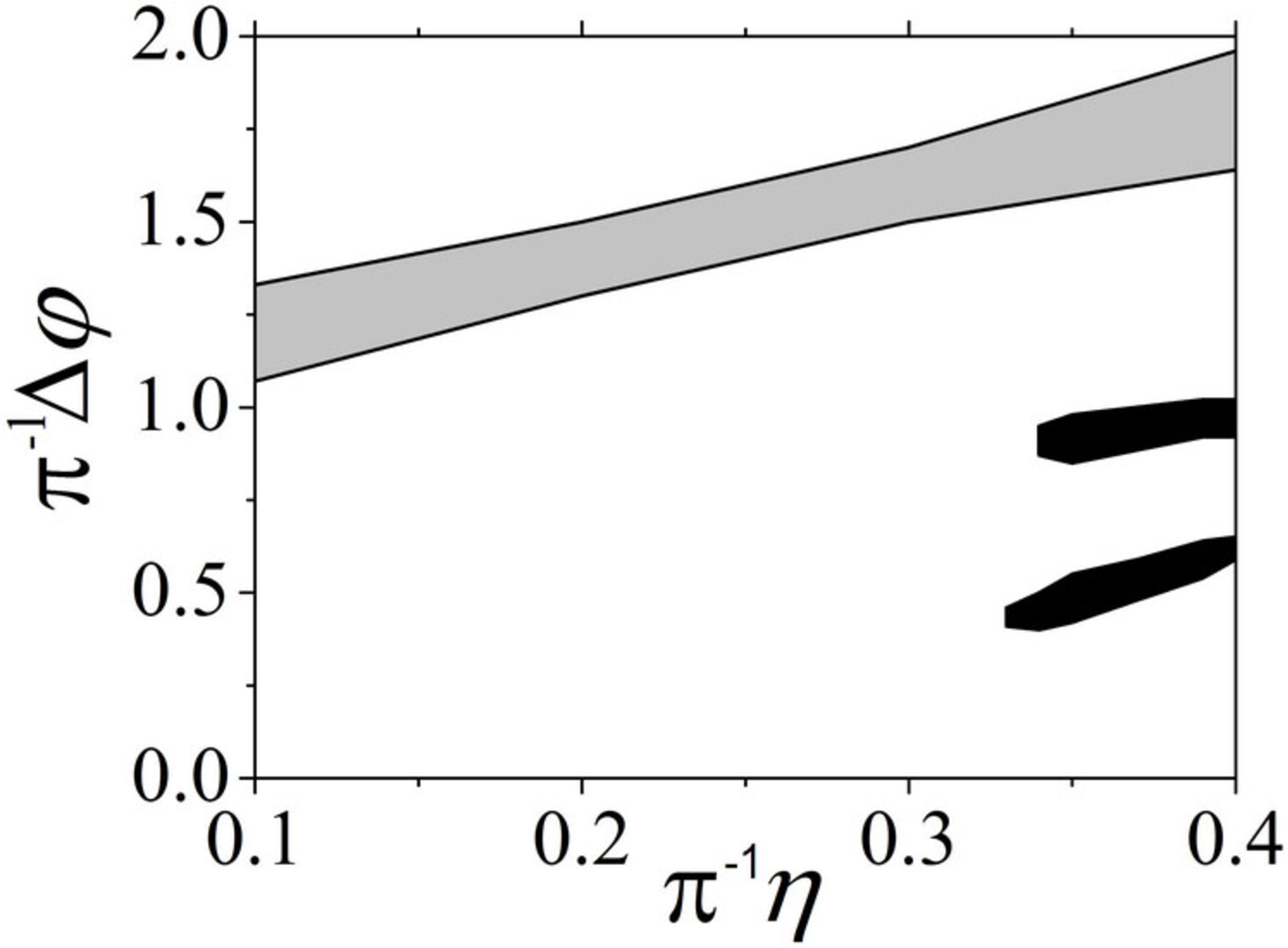}}
\caption{(Color online) (a) The plane of $(\protect\eta ,\Delta \protect%
\varphi )$ (with $\protect\eta \in \left[ 0.1\protect\pi ,0.4\protect\pi %
\right] $) for outcomes of collisions between lighter solitons, with $P=1.3$%
. The merger and rebound occur in the white and light-gray areas,
respectively (the symmetric merger is observed along the dashed lines). The
colliding solitons destroy each other in the dark-gray area. Points A $(%
\protect\eta ,\Delta \protect\varphi )=(0.35,1.7)\protect\pi $, B $(\protect%
\eta ,\Delta \protect\varphi )=(0.35,0.7)\protect\pi $, C $(\protect\eta %
,\Delta \protect\varphi )=(0.35,0.25)\protect\pi $, and D $(\protect\eta %
,\Delta \protect\varphi )=(0.35,0)\protect\pi $ represent typical examples
of the full rebound, destruction, symmetric merger and asymmetric merger,
respectively. (b) The same plane for collisions between heavier solitons,
with $P=1.8$. The white and light-gray areas have the same meaning as in
(a). Quasi-elastic collisions occur in the black regions.}
\label{etaplane}
\end{figure}

\begin{figure}[tbp]
\centering%
\subfigure[] {\label{fig10a}
\includegraphics[scale=0.33]{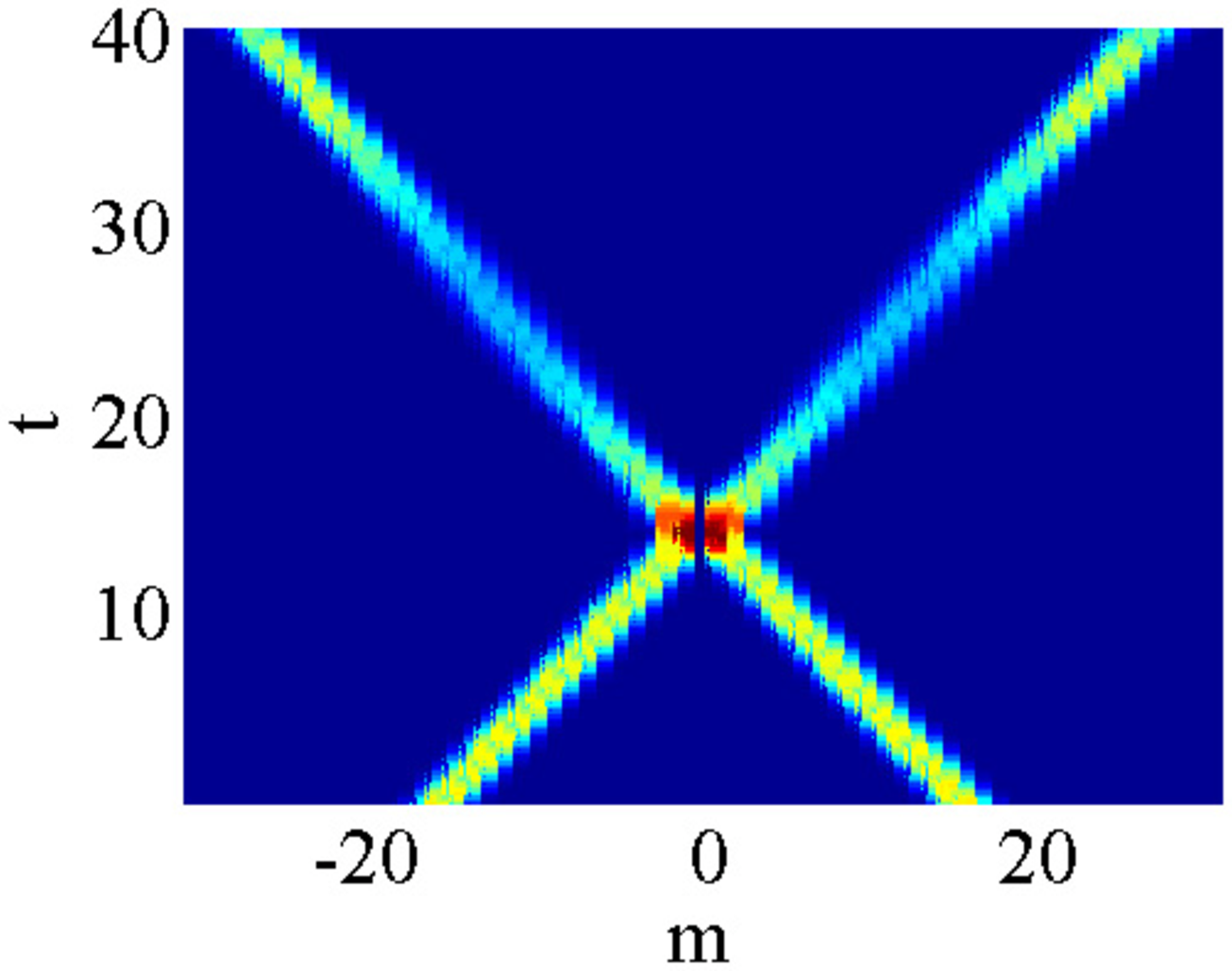}}%
\subfigure[] {\label{fig10b}
\includegraphics[scale=0.33]{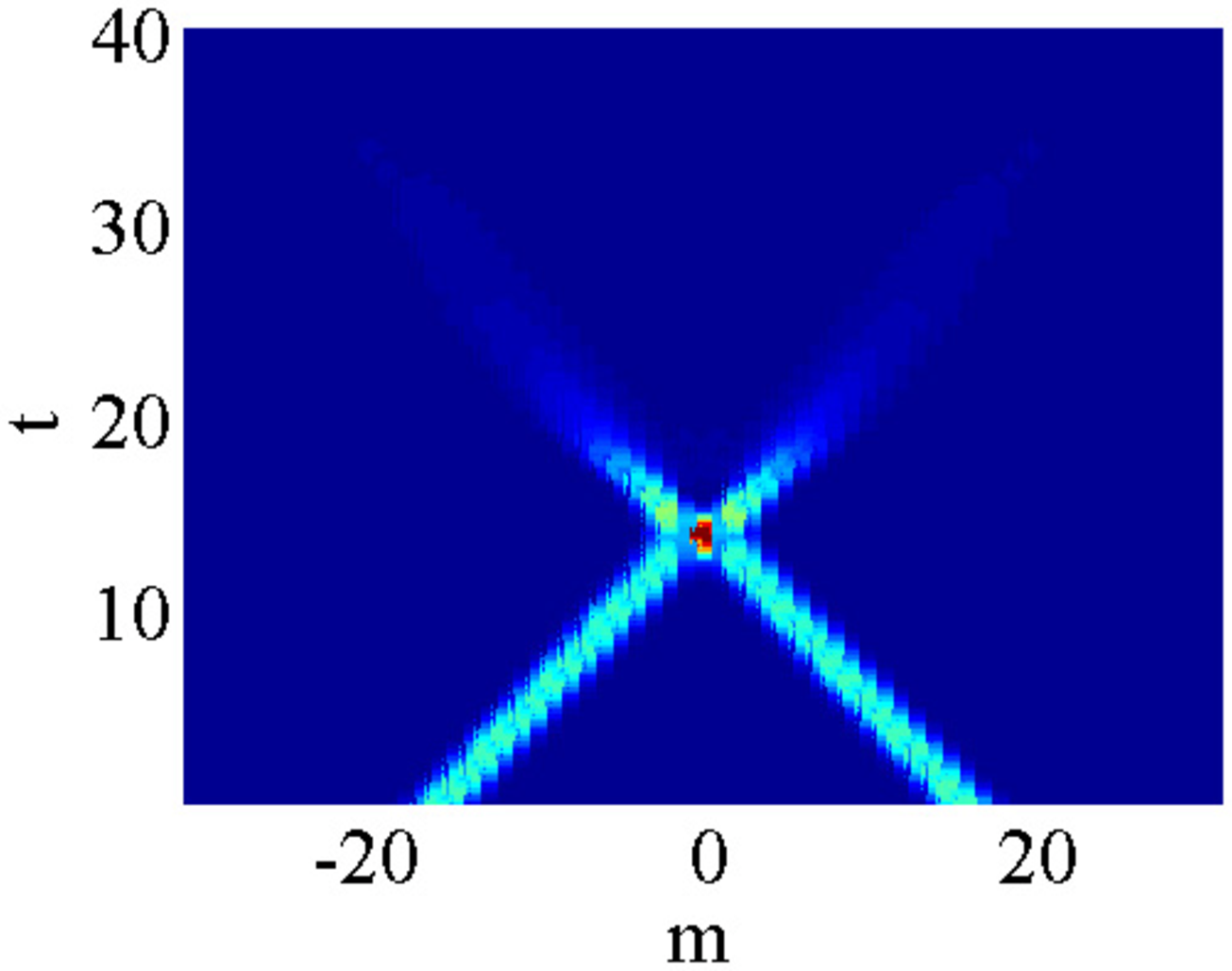}}
\subfigure[]{\label{fig10c}
\includegraphics[scale=0.33]{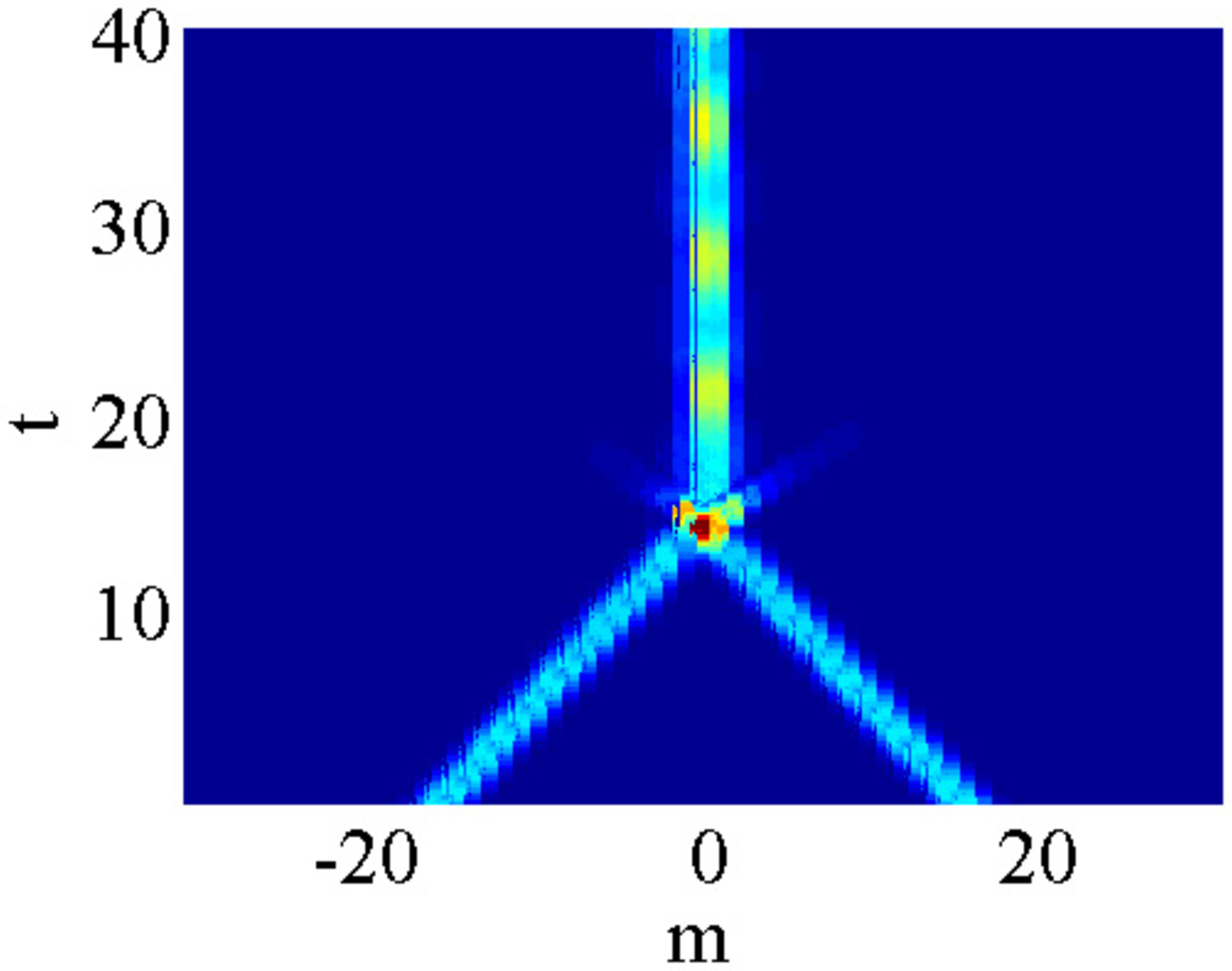}}
\subfigure[]{\label{fig10d}
\includegraphics[scale=0.33]{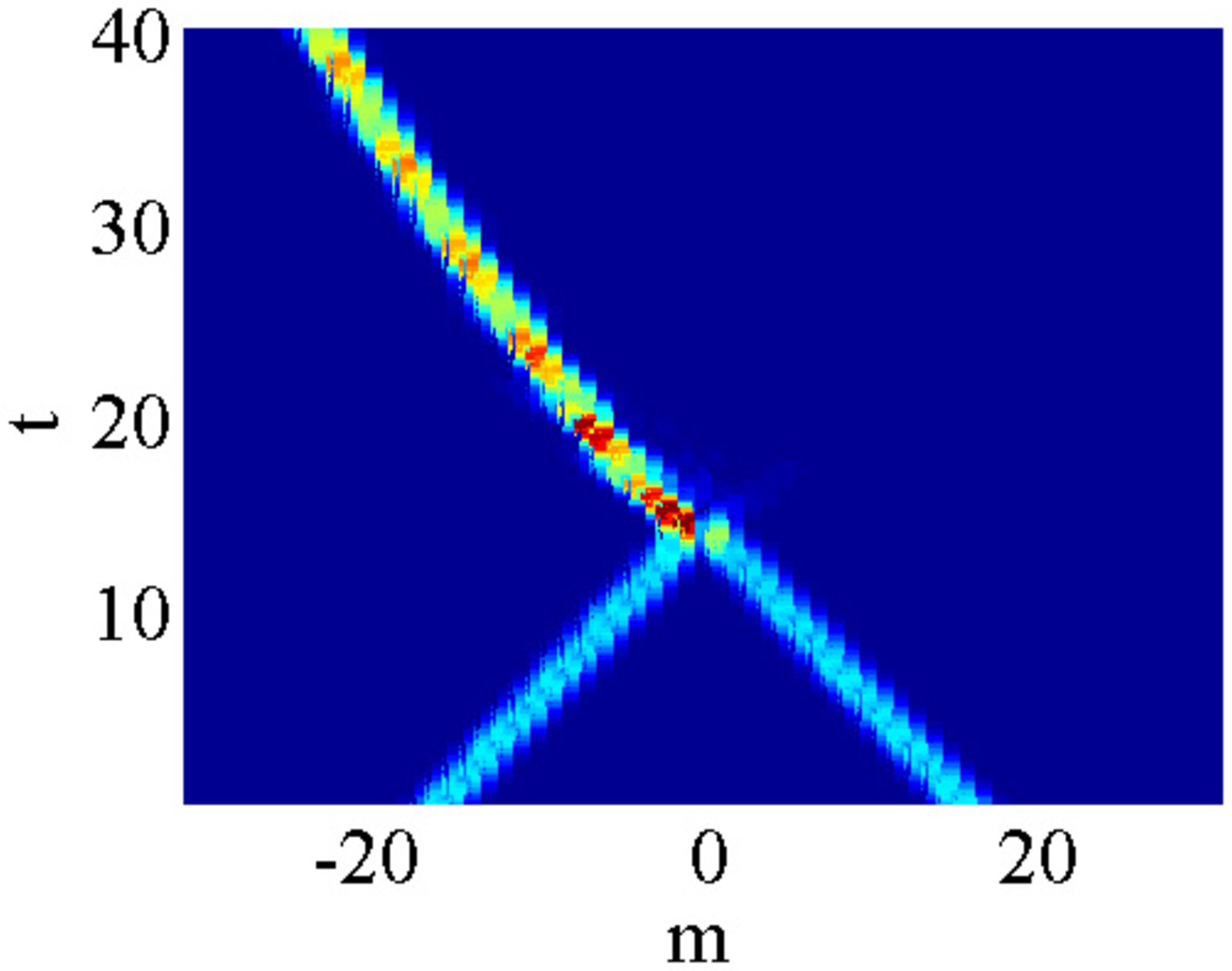}}
\subfigure[]{\label{fig10e}
\includegraphics[scale=0.33]{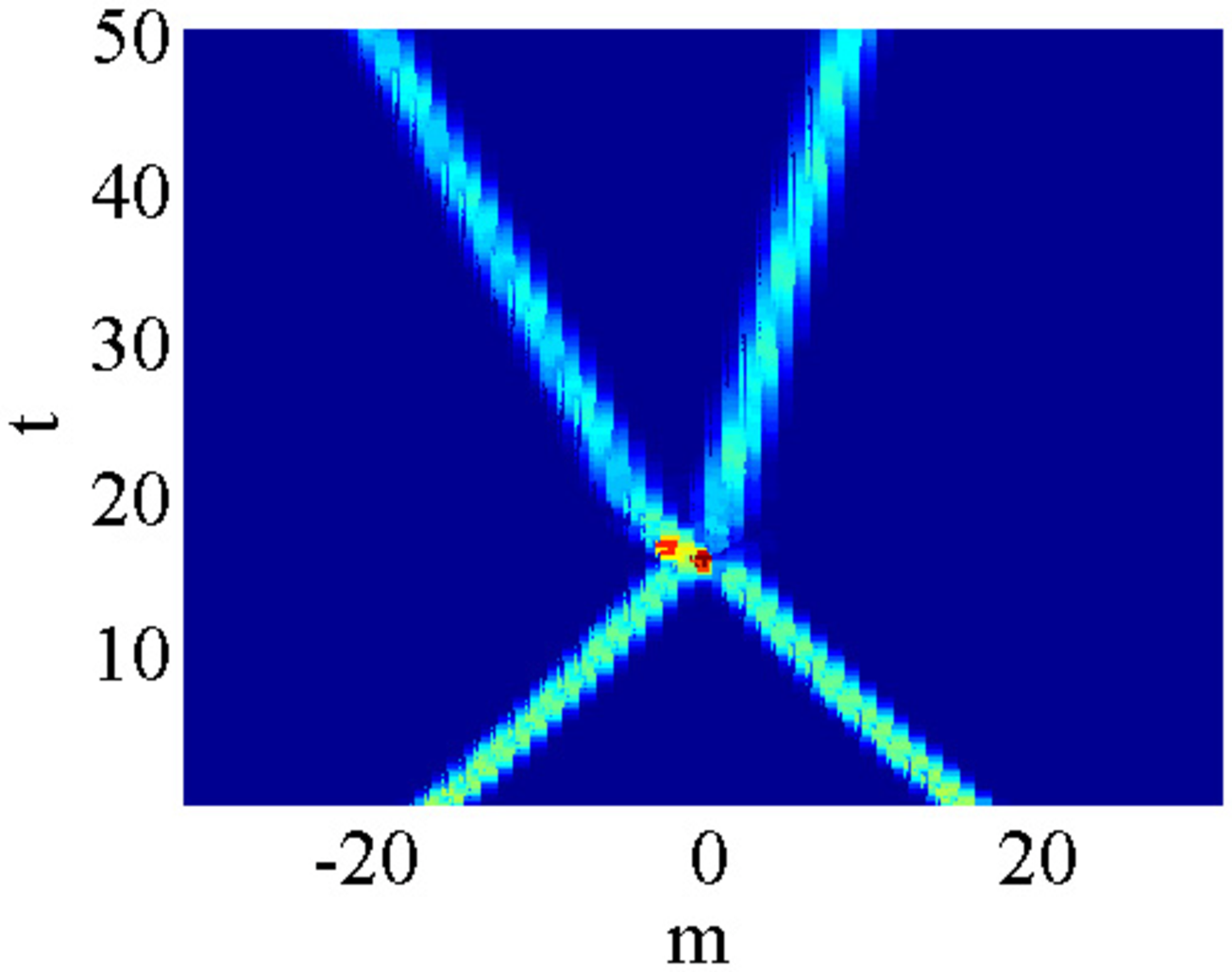}}
\subfigure[]{\label{fig10f}
\includegraphics[scale=0.33]{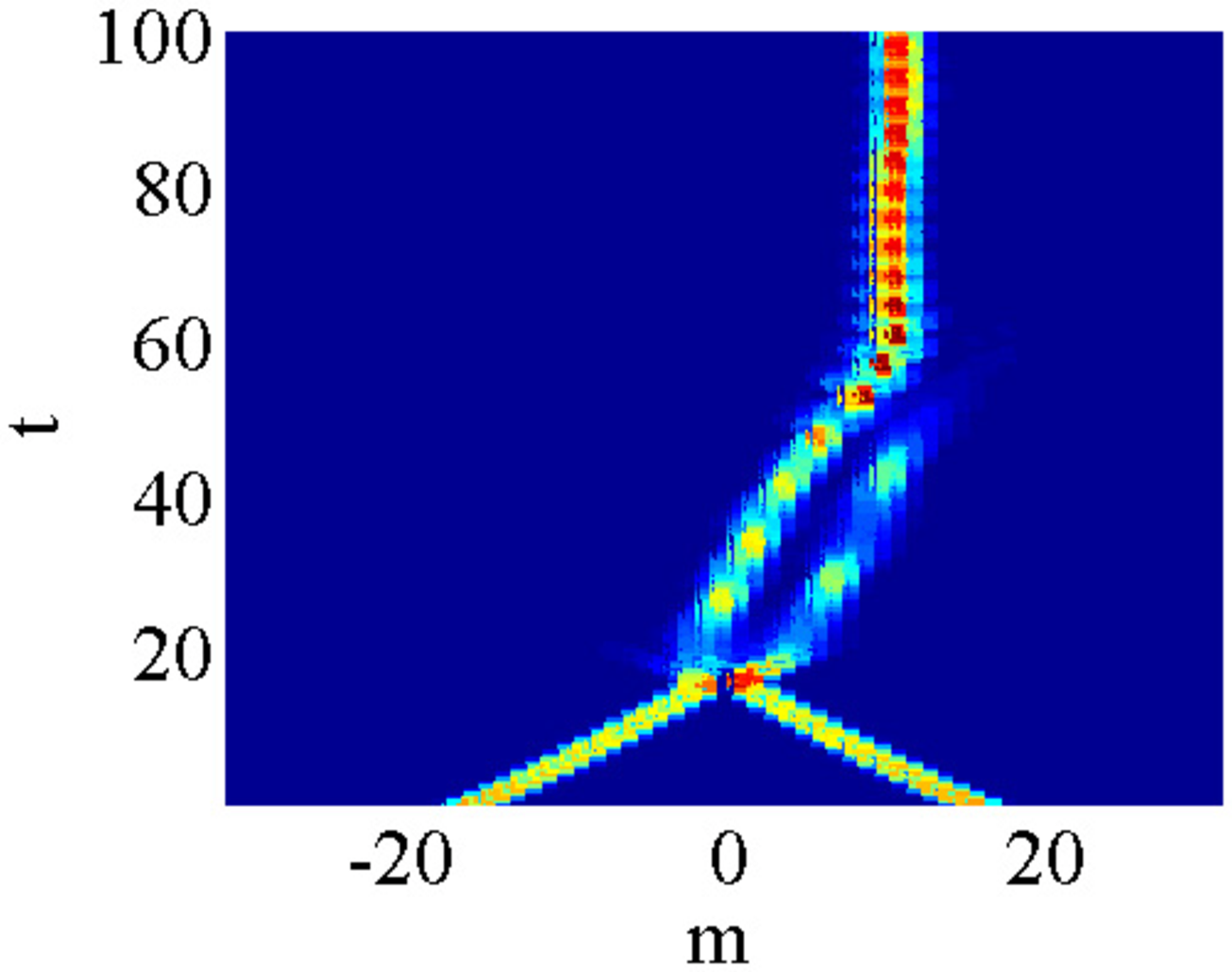}}
\caption{(Color online) (a)-(d) Typical examples of collisions between
solitons with norms $P=1.3$ (shown in the cross section $n=0$),
corresponding to points A-D in Fig. \protect\ref{fig9a}. (e) An example of
the quasi-elastic collision between heavier solitons, with $P=1.8$, and $(%
\protect\eta ,\Delta \protect\varphi )=(0.35,0.95)\protect\pi $. (f) An
example of the multiple-rebound collision ending up with merger, for $P=1.8$
and $(\protect\eta ,\Delta \protect\varphi )=(0.35,1.58)\protect\pi $.}
\label{Movingexample}
\end{figure}

\subsection{Collision between moving solitons}

The robust mobility of the 2D discrete solitons in the horizontal direction
suggests to study collisions between them \cite{Papa}-\cite{Meier}. To this
end, we simulated Eq. (\ref{discr}) with the input composed of two solitons
separated by distance $2m_{0}$, i.e., with their centers placed at sites $%
\left( \mp m_{0},0\right) $, and phase shift $\Delta \varphi $, which is
another control parameter of the collision, in addition to the relative
velocity determined by kicks $\pm \eta $ \cite{Papa}-\cite{Khaykovich37}:
\begin{equation}
\psi _{mn}(t=0)=\phi _{-m_{0},0}e^{i\eta m}+\phi _{m_{0},0}e^{-i\eta
m+\Delta \varphi },  \label{colli_initial}
\end{equation}%
In the simulations, the size of the domain was $64\times 64$, with $m_{0}=16$%
. The norm of each soliton was $P=1.3$ or $1.8$, to explore the collision of
relatively light and heavy solitons, respectively.

For $P=1.3$, the moving solitons feature three major types of collisions in
the plane of $\left( \eta ,\Delta \varphi \right) $, as shown in Fig. \ref%
{fig9a}: rebound, destruction, and merger. Particular examples of the
rebound, destruction, symmetric merger, and asymmetric merger are displayed
in Figs. \ref{fig10a}-\ref{fig10d}, which correspond, respectively, to
sample points marked in Fig. \ref{fig9a}: A $(\eta ,\Delta \varphi
)=(0.35,1.7)\pi $, B $(\eta ,\Delta \varphi )=(0.35,0.7)\pi $, C $(\eta
,\Delta \varphi )=(0.35,0.25)\pi $ and D $(\eta ,\Delta \varphi
)=(0.35,0)\pi $. Results of the direct simulations are displayed in cross
section $n=0$.

At larger values of the norm (here, at $P=1.8$), the moving solitons are
more robust with respect to collisions. Namely, in this case destruction is
not observed in the considered interval of the values of the kick, $\eta \in
\lbrack 0.1\pi ,0.4\pi ]$, while a new outcome occurs, in the form of
quasi-elastic collisions (see Fig. \ref{fig10e}). Parametric regions of
three types of outcomes of collisions between heavier solitons, \textit{viz}%
., the merger, rebound, and quasi-elastic interaction, are shown in Fig. \ref%
{fig9b}, in the same plane as in Fig. \ref{fig9a}. Because the merger and
the rebound are quite similar to what was shown above for $P=1.3$, here, in
Fig. \ref{fig10e}, we display only a typical example of the quasi-elastic
collision. It is also worthy to note that, in a parametric area close to the
rebound [i.e., in the light gray area in Fig. \ref{fig9b}], colliding
solitons merge after multiple rebounds, see a typical example in Fig. \ref%
{fig10f}.

\begin{figure}[tbp]
\centering%
\subfigure[] {\label{fig11a}
\includegraphics[scale=0.33]{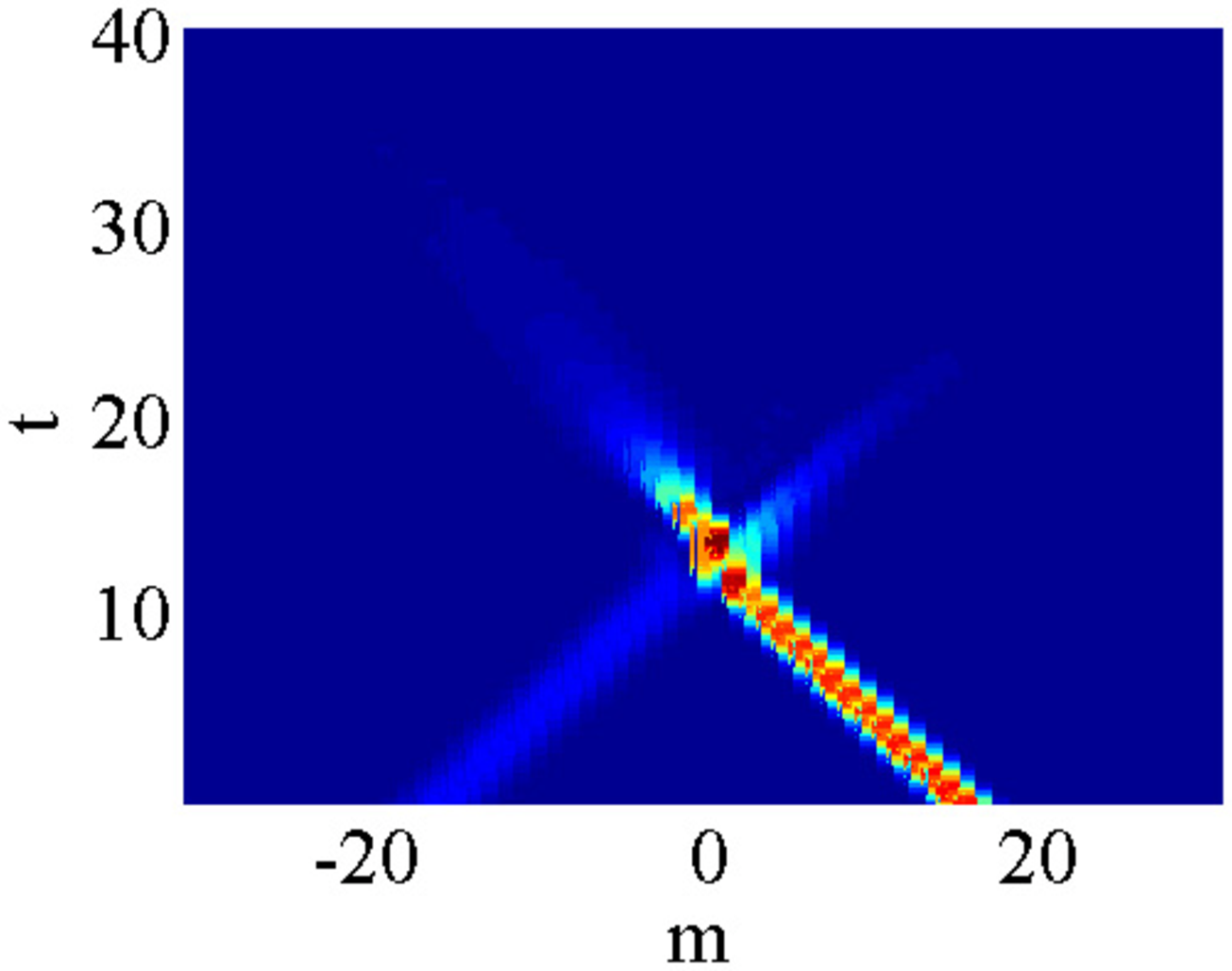}}%
\subfigure[] {\label{fig11b}
\includegraphics[scale=0.33]{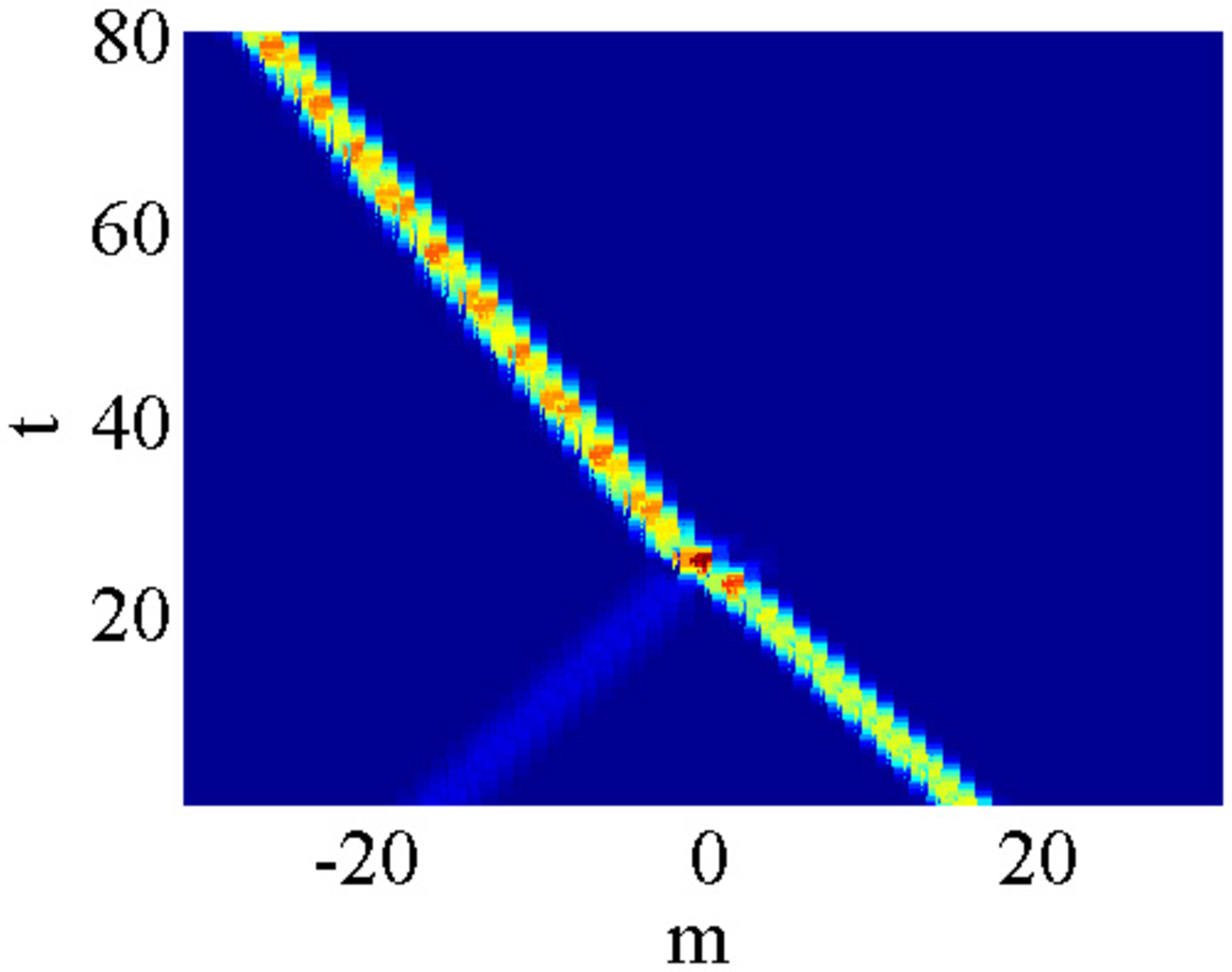}}
\subfigure[]{\label{fig11c}
\includegraphics[scale=0.33]{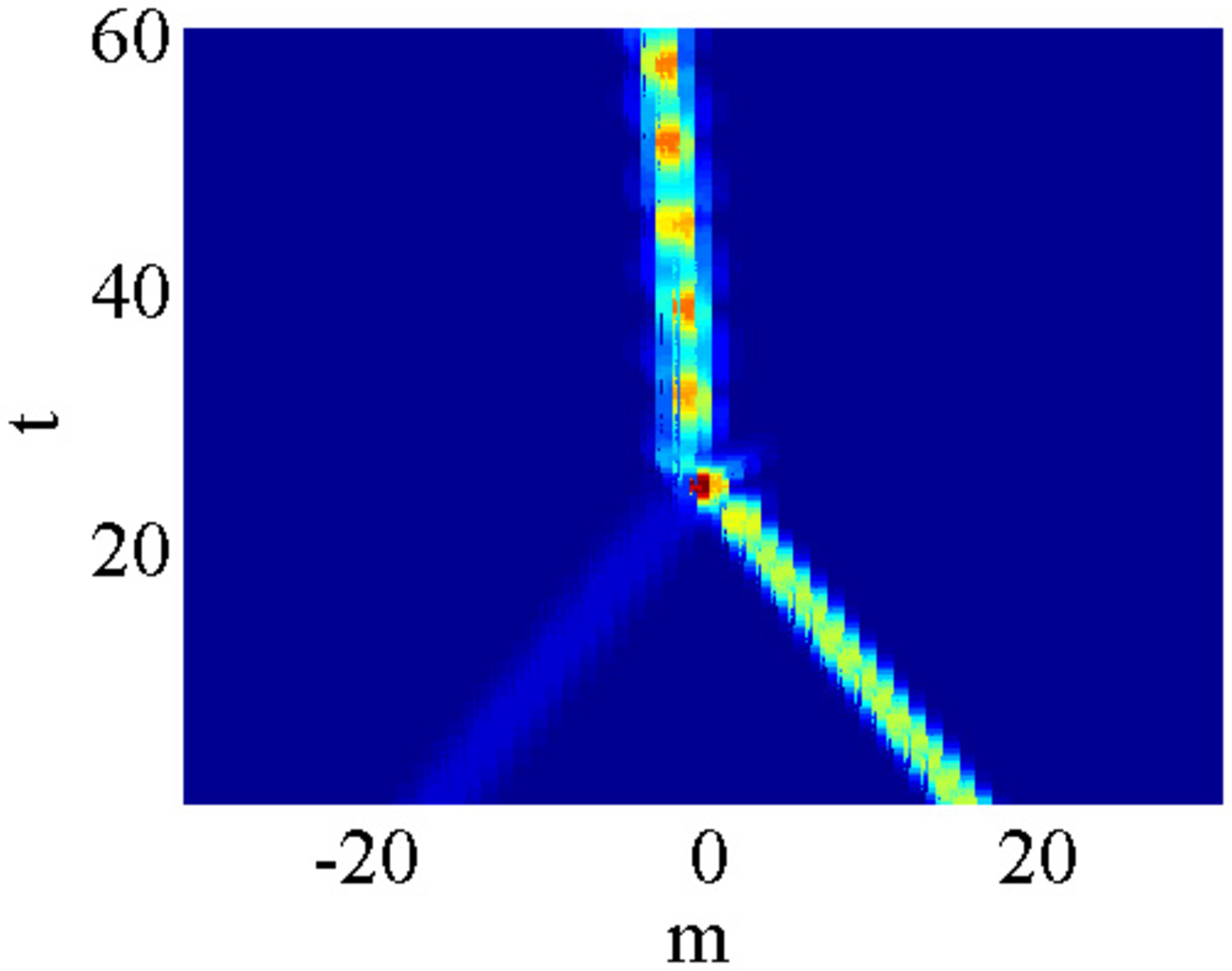}}
\caption{(Color online) Outcomes of collisions between solitons with unequal
total powers, $P_{1}=1.0$ and $P_{2}=1.5$: (a) destruction, for kicks $%
\protect\eta =\pm 0.40\protect\pi $ and phase shift $\Delta \protect\varphi %
=0$; (b) merger, for $\protect\eta =0.20\protect\pi $ and $\Delta \protect%
\varphi =0$; (c) also merger, for $\protect\eta =0.20\protect\pi $ and $%
\Delta \protect\varphi =1.42\protect\pi $.}
\label{fig11}
\end{figure}

Imbalance between powers (norms) of the colliding solitons, $P_{1}\neq P_{2}$%
, also affects outcomes of the collisions, cf. Ref. \cite{OL1996}. For
instance, the asymmetric merger of two solitons with equal powers, $%
P_{1}=P_{2}=1.3$, which collide under the action of kicks $\eta =\pm 0.35\pi
$ with the zero phase shift, $\Delta \varphi =0$, displayed in Fig. \ref%
{fig10d}, is replaced by mutual destruction for unequal powers $P_{1}=1.0$, $%
P_{2}=1.5$ (chosen so that the net power, $P_{1}+P_{2}$, remains nearly the
same as before), for nearly the same values of the kicks, $\eta =\pm 0.40\pi
$, and again with $\Delta \varphi =0$, see Fig. \ref{fig11a}. However, the
reduction of the kick to $\eta =\pm 0.20\pi $ switches the outcome of the
collision for the same power-imbalance soliton pair from the destruction to
merger into a single moving soliton, as shown in Fig. \ref{fig11b}, which is
quite similar to the merger observed in Fig. \ref{fig10d}. On the other
hand, the symmetric collision leading to the merger of the solitons with $%
P_{1}=P_{2}=1.3$ into a single standing soliton, for $\eta =\pm 0.35$ and $%
\Delta \varphi =0.20\pi $ in Fig. \ref{fig10c}, extends, for $P_{1}=1.0$, $%
P_{2}=1.5$ and $\eta =\pm 0.20\pi $, $\Delta \varphi =1.42\pi $, into a
similar merger, although this time the emerging single soliton is moving,
see Fig. \ref{fig11c}.

\section{Conclusions}

The objective of this work is to study 2D matter-wave solitons in the
ultracold gas trapped in a deep OL (optical-lattice) potential, with
quadrupole-quadrupole interactions (QQIs) between particles. The quadrupoles
are introduced as tight bound states of dipoles and anti-dipoles, whose
polarizations are orientated in the $z$-direction, i.e., perpendicular to
the 2D plane in which the gas is trapped. The polarization of the
quadrupoles is imposed by the dc electric field, which is also directed
along $z$, but with the strength linearly growing along the $x$-direction.
Such a field configuration, which may be created by means of a tapered
capacitor, is necessary because the quadrupoles interact locally with the
gradient of the external field. The setting is described by the discrete
(lattice) Gross-Pitaevskii equation, due to the fragmentation of the
condensate by the strong OL. Together with the long-range QQI between
lattice sites, the contact interaction, that may be either attractive or
repulsive, is taken into account as the onsite nonlinearity in the resulting
lattice model.

The shapes, stability, mobility and collisions of 2D fundamental lattice
solitons were studied by means of numerical simulations. It has been found
that stable solitons exist above a threshold value of the norm, which is
lower than for matter-wave solitons in dipolar lattice-trapped gases. The
threshold is much weaker affected by the contact interactions than in the
case of the DDI (dipole-dipole long-range interaction). The shape of the
solitons features anisotropy, which is stronger for heavier solitons. The
mobility of the discrete solitons on 2D lattices with long-range intersite
interactions is studied here for the first time, also exhibiting strong
anisotropy. Collision between mobile solitons were explored too. Collisions
between lighter solitons may lead to their merger, rebound and destruction.
Heavier solitons are not destroyed by collisions. Estimates of physical
parameters demonstrate that the proposed setting may be experimentally
implemented in gases of small molecules, metastable alkaline-earth metals,
or alkaline diatoms.

A natural extension of the work may be to construct bound states of 2D
lattice solitons. A challenging possibility is to look for topological 2D
solitons, such as lattice vortices, in this discrete anisotropic system.

\begin{acknowledgments}
We appreciate valuable discussions with A. Maluckov. This work was supported
by Chinese agency CNNSF (grants No. 11104083, 11204089, 11205063), by the
German-Israel Foundation through grant No. I-1024-2.7/2009, and by the Tel
Aviv University in the framework of the ``matching" scheme for a
postdoctoral fellowship of Y.L.
\end{acknowledgments}

\bibliographystyle{plain}
\bibliography{apssamp}

\end{document}